\documentclass[bimj,fleqn]{w-art}
\usepackage{times}
\usepackage{w-thm}
\usepackage[authoryear]{natbib}
\theoremstyle{plain}

\theoremstyle{definition}

\usepackage[]{graphicx}
\usepackage{moreverb,url}
\usepackage[colorlinks,bookmarksopen,bookmarksnumbered,citecolor=red,urlcolor=red]{hyperref}
\usepackage{natbib,graphicx,setspace,amsmath,amssymb,subfigure,url,multirow,booktabs,verbatim,bm,epstopdf,url}

\chardef\bslash=`\\ 

\begin{document}
\DOIsuffix{bimj.200100000}
\Volume{52}
\Issue{61}
\Year{2010}
\pagespan{1}{}
\keywords{Censoring; Dynamic prediction; Joint model; Longitudinal data; Reduced rank functional principle component model\\[1pc]
\noindent\hspace*{-4.2pc} Supporting Information for this article is available from the author or on the WWW under\break \hspace*{-4pc} \underline{http://dx.doi.org/10.1022/bimj.XXXXXXX}
}  

\title[A Flexible Joint Model for  Multiple Longitudinal Biomarkers and A Time-to-Event Outcome]{A Flexible Joint Model for  Multiple Longitudinal Biomarkers and A Time-to-Event Outcome: With Applications to Dynamic Prediction Using Highly Correlated Biomarkers}
\author[Ning Li {\it{et al.}}]{Ning Li\inst{1}}
\address[\inst{1}]{Departments of Medicine and Biomathematics, University of California at Los Angeles, Los Angeles, California 90095, U.S.A.}
\author[]{Yi Liu\inst{2}}
\address[\inst{2}]{School of Mathematical Sciences, Ocean University of China, Qingdao 266100, China}
\author[]{Shanpeng Li\inst{3} }
\address[\inst{3}]{Department of Biostatistics, University of California at Los Angeles, Los Angeles, California 90095, U.S.A.}
\author[]{Robert M. Elashoff\inst{4} }
\address[\inst{4}]{Department of Biomathematics, University of California at Los Angeles, Los Angeles, California 90095, U.S.A.}
\author[]{Gang Li\footnote{Corresponding author: {\sf{e-mail: vli@ucla.edu}}, Phone: +00-310-206-5865, Fax: +00-310-267-2113}\inst{3} }
\Receiveddate{zzz} \Reviseddate{zzz} \Accepteddate{zzz}

\begin{abstract}
In biomedical studies it is common to collect data on multiple biomarkers during study follow-up
for dynamic prediction of a time-to-event clinical outcome. The biomarkers are typically intermittently measured, missing
at some  event times, and may be subject to high biological variations, which cannot be readily used as time-dependent covariates in a standard time-to-event model. Moreover, they can be highly correlated if they are from in the same biological pathway. To address these issues, we propose a flexible joint model framework that models the multiple biomarkers with a  shared   latent reduced rank longitudinal principal component model   and correlates the latent process to the event time by the Cox model for dynamic prediction of the event time. The proposed joint model for highly correlated biomarkers is more flexible than some existing  methods since the latent trajectory shared by the multiple biomarkers does not require specification of a priori parametric time trend and is determined by data. We derive an Expectation-Maximization (EM) algorithm for parameter estimation,  study large sample properties of the estimators, and adapt the developed method to make dynamic prediction of the time-to-event outcome. Bootstrap is used for standard error estimation and inference. The proposed method  is evaluated using simulations and illustrated on a lung transplant data to predict chronic lung allograft dysfunction (CLAD) using chemokines measured in bronchoalveolar lavage fluid of the patients.
\end{abstract}

\maketitle                   






\section{Introduction}

In clinical studies, researchers are often interested in monitoring certain important clinical events and predicting the events using repeatedly measured response variables or biomarkers during follow-up. As an example we consider a lung transplant study in which multiple chemokines (biomarkers) in bronchoalveolar lavage fluid of the patients were measured repeatedly post transplant. Of research interest is to evaluate biomarker predictiveness for some important clinical events such as chronic lung allograft dysfunction (CLAD). A conventional statistical approach is to fit a Cox regression model in which those biomarkers are included as time-dependent covariates. However, such a model is regarded inapplicable if the biomarkers are
only measured intermittently,  missing at certain event times, or subject to measurement error \citep{Prentice:1982}.  Furthermore, some biomarkers are highly correlated because they belong to the same biological pathway. Including them directly into a regression model could suffer the collinearity problem and unstable model fit.

A popular approach to addressing the above issues in dynamic prediction is to model the  biomarker(s) using a shared latent process and then incorporate the latent process into a survival model for the event of interest.
A rich set of joint models have been developed using this approach for a single  biomarker; see, among others,
\citet{Wulfsohn:1997, Elashoff:2016,Song:2002, Wang:2001} and the references therein.
Some joint models using this approach have also been developed for multiple biomarkers and time-to-event data, where the association among multiple endpoints is modeled through either latent class \citep{Proust:2009a} or latent random variables \citep{Huang:2001,Chi:2006,Hatfield:2011,Rizopoulos:2011}. These models assume that the multiple longitudinal outcomes exhibit correlated, but also to some extent, distinct patterns of trajectories so that their evolutionary paths are characterized separately. More recently, \citet{Luo:2014} and \citet{He:2016} introduced a multilevel model in which the longitudinal outcomes share a latent disease severity process that is expressed as a linear function of fixed and random effects, and an accelerated failure time model or Cox model is used for the event time. Their model allows for highly correlated biomarkers, but requires specification of a priori parametric time trend for the latent process, which is not always an easy task in practice.

{
The purpose of this paper is to study a flexible joint model framework where the multiple biomarkers for each subject are modeled using a shared  latent reduced rank longitudinal principal component model  and  the latent process is correlated to the event time through a Cox model for dynamic prediction of the event. No priori parametric time trend assumption is required for the latent process in our model.
As described later in equation (\ref{long2}) of Section 2,  each individual latent trajectory in our model is expressed as an additive function of the overall mean curve and several principal component curves, with the latter describing the modes of variation in individual trajectories. The principal component functions are estimated directly and weighted by mutually independent random effects. Compared to the traditional mixed effects method for functional principal component analysis, the reduced rank method estimates fewer parameters, so the fitted curve is generally more stable and accurate. In our joint model, the latent process estimation is data-driven and non-parametric in the sense that the overall mean and principal component functions are modeled by B-splines, which is a flexible approach to complement the models of \citet{Luo:2014} and \citet{He:2016}. We point out that our model for the latent process is inspired by  the reduced rank mixed effects framework for principal component analysis of functional data developed by \citet{James:2000}, which is well suited for situations with irregular and sparse observation time points across individuals
as exhibited in the above mentioned lung transplant study. The reduced rank functional principle component approach has also been previously adopted by \citet{Yao:2007} to jointly model a single biomarker and an event time \cite{Yao:2007}. Our work in this paper is a natural, but nontrivial extension of
\citet{Yao:2007} from the single biomarker case to the multiple biomarkers case with some further developments. Specifically, in addition to extending the estimation procedure to the multiple biomarker setting, we develop the large sample theory for the resulting estimator which has not been investigated previously. We further propose a dynamic prediction tool which has high practical relevance for dynamic prediction of a clinical outcome from multiple highly correlated  biomarkers. Lastly, we have implemented the proposed joint model in an R package \textbf{\texttt{JMM}} which is available for download at \underline{https://github.com/shanpengli/JMM.}

This article is organized as follows. Details of the joint model formulation are given in Sections 2.1. The maximum likelihood estimators and their asymptotic properties are given in Section 2.2. Section 2.3 outlines the derivation of a predictive accuracy measure. Application of this joint model to the lung transplant study is given in Section 3. Section 4 presents some simulation results to evaluate the
proposed method. Section 5 gives some concluding remarks.

\section{MODEL AND ESTIMATION METHODS} \label{jointmodel}
\subsection{The Joint Model} \label{modelsetup}

Let $Y_{ij}(t)$ be the $j^{th}$ biomarker measured on subject $i$ at time $t$, $i = 1, \ldots, n$, $j = 1, \ldots, J$, and $t \in [0, \tau]$ for some known $\tau > 0$. We employ a latent process approach to model the inter-correlation among the $J$ biomarkers and assume that they share a common subject-specific stochastic trend over time, namely $\mu_i(t)$. This assumption is realistic if the biomarkers are in the same biological pathway and exhibit similar trajectories in preliminary data analysis. Specifically, the $j^{th}$ biomarker, $j = 1, \ldots, J$, is assumed to be
\begin{eqnarray}
Y_{ij}(t)=X_{ij}(t)^T\beta_{0j}+\mu_i(t)\beta_{1j}+\epsilon_{ij}(t),
\label{long1}
\end{eqnarray}
where $X_{ij}(t)$ is a $p_j \times 1$ vector of possibly time-dependent covariates, $\beta_{0j}$ the associated regression coefficients, $\mu_i(t)$ the latent process at time $t$, $\beta_{1j}$ the factor loading of $\mu_i(t)$ for the $j^{th}$ outcome, and $\epsilon_{ij}(t) \sim i.i.d. N(0, \sigma_j^2)$ the measurement error. Note that $\beta_{11}$ is set to 1 for the purpose of identifiability. Write $\beta_{j} = (\beta_{0j}, \beta_{1j})$. We further assume the biomarkers are associated with the event risk through the latent process $\mu_i(t)$ which characterizes the overall underlying trend shared by these biomarkers.

Following the reduced rank mixed effects model by \citet{James:2000}, the latent process $\mu_i(t)$ in Equation (\ref{long1}) is characterized by a additive function of mean curve $\mu(t)$ and principal component curves $f_{\kappa}(t)$, $\kappa = 1, \ldots, k$; the latter capture principal patterns of individual variation around the mean curve. Specifically,
\begin{eqnarray}
\mu_i(t)=\mu(t) + \sum_{\kappa = 1}^k f_{\kappa}(t)\alpha_{i\kappa} = b(t)^T\theta+b(t)^T\Theta\alpha_i,
\label{long2}
\end{eqnarray}
where $\mu(t)$ and $f_{\kappa}(t)$ are flexibly modeled by $q$-dimensional smooth basis spline functions $b(t)$, $\theta$ and $\Theta$ are $q \times 1$ vector and $q \times k$ matrix of spline coefficients, respectively, subject to $\Theta^T\Theta = I$, and $\int b(t)b(t)^\top dt = I$, to impose orthogonality constraints on the principal component functions. The random effects $\alpha_i = (\alpha_{i1}, \ldots, \alpha_{ik})^T$ represent individual variation in the relative weights of the principal component functions across study subjects, and are assumed to be $N_k(0, D)$ with $D$ being a $k \times k$ diagonal matrix to avoid confounding with $\Theta$. Models (\ref{long1}) and (\ref{long2}) indicate that the data correlations across time and between multiple biomarkers are characterized by the random effects $\alpha_i$.

The hazard function of event times is assumed to take the form of a Cox regression:
\begin{eqnarray}
\lambda_i(t) = \lambda(t)\exp(Z_i(t)^T\eta+\gamma\mu_i(t)),
\label{surv}
\end{eqnarray}
where the baseline hazard function $\lambda(t)$ is completely unspecified and $Z_i(t)$ is a vector of possibly time-dependent covariates with unknown coefficients $\eta$. The direction and magnitude of the association between the biomarker latent process $\mu_i(t)$ and the event risk is characterized by the parameter $\gamma$. Thus, testing $\gamma = 0$ is equivalent to testing the association between longitudinal and survival endpoints.

In real practice, selection of the number of knots in the spline basis can be based on cross-validated loglikelihood. However, given the computational burden of cross-validation, \citet{Rice:2001} proposed the use of AIC or BIC to reduce computation cost; comparable results were obtained by AIC, BIC and cross-validation in the datasets they examined.

To determine the optimal rank $k$ of principal component functions, \citet{James:2000} suggested two approaches. The first approach is to calculate the proportion of total variation explained by each principal component, which can be approximated by $D_{\kappa,\kappa}/tr(D)$ where $D_{\kappa,\kappa}$ is the variance of the random effect that is associated with the $\kappa$th principal component function and $tr(D)$ is the trace of the covariance matrix $D$. This quantity can be used to examine if including an additional principal component only explains a small amount of total variation given the other principal components that are already in the model. The second approach is to keep track of increment in loglikelihood as $k$ increases, and choose the optimal rank where the increase in loglikelihood levels off.

\subsection{Estimation and Inference}

Let $Y_i$ be the longitudinal measurements of all biomarkers and $T_i$ the actual event time for subject $i$ that may be censored by $C_i$, $i = 1, \ldots, n$, so we observe $\tilde{T}_i = min (T_i, C_i)$ and $\Delta_i = I(T_i \leq C_i)$. Noninformative censorship is assumed here; that is, $T_i \perp C_i$. An underlying assumption in the joint model specified in (\ref{long1}), (\ref{long2}), and (\ref{surv}) is that $Y_i$ and $T_i$ are conditionally independent given the random effects $\alpha_i$ and covariates $X_i$ and $Z_i$. The observed likelihood function is thus
\begin{eqnarray}
L(\psi; Y, \tilde{T}, \Delta)&=&\prod_{i=1}^n f(Y_i,\tilde{T}_i,\Delta_i|\psi) \nonumber\\
&=&\prod_{i=1}^n\int f(Y_i|\alpha_i,\psi)f(\tilde{T}_i,\Delta_i|\alpha_i,\psi)f(\alpha_i|\psi)d\alpha_i \nonumber \\
&=&\prod_{i=1}^n\int f(Y_i|\alpha_i,\psi)[f(\tilde{T}_i|\alpha_i,\psi)^{\Delta_i}(1-F(\tilde{T}_i|\alpha_i,\psi))^{1-\Delta_i}] f(\alpha_i|\psi)d\alpha_i,
\label{likelihood}
\end{eqnarray}
where $\psi$ collects all the model parameters. The complete-data likelihood function $L(\psi; Y, \tilde{T}, \Delta, \alpha)$ is defined based on~(\ref{likelihood}) assuming the random effects $\alpha_i$ are known.

Write $\psi = (\phi, \Lambda)$, where $\Lambda(t) = \int_0^t \lambda(u)du$ and $\phi$ contains the remaining parameters. The maximum likelihood estimate $\hat{\psi}$ maximizes the likelihood over a space in which $\phi$ belongs to a bounded set and $\Lambda$ is an increasing functions in $t$ with $\Lambda(0) = 0$. The likelihood function is difficult to maximize directly in the presence of integrals, so we consider to obtain $\hat{\psi}$ through an expectation-maximization (EM) algorithm which iterates between an \textit{Expectation} step (E-step) and a \textit{Maximization} step (M-step) \citep{Dempster:1977,Elashoff:2008}. In the E-step we compute the expected value of complete-data log-likelihood with respect to $\alpha_i$ conditional on $(Y_i, \tilde{T}_i, \Delta_i, \psi^{(m)})$, where $\psi^{(m)}$ is the current estimate of $\psi$ and $m = 0, 1, \ldots$ denotes iterations. Specifically, for any function $h(\cdot)$ of $\alpha_i$ that appears in the complete-data log-likelihood, its expectation can be evaluated by
\begin{eqnarray}
E\{h(\alpha_i)|Y_i,\tilde{T}_i,\Delta_i,\psi^{(m)}\}&=&\int h(\alpha_i)f(\alpha_i|Y_i,\tilde{T}_i,\Delta_i,\psi^{(m)})d\alpha_i \nonumber \\
&=&\frac{\int h(\alpha_i)f(\alpha_i,Y_i,\tilde{T}_i,\Delta_i|\psi^{(m)})d\alpha_i}{f(Y_i,\tilde{T}_i,\Delta_i|\psi^{(m)})} \nonumber \\
&=&\frac{\int h(\alpha_i)f(Y_i,\tilde{T}_i,\Delta_i|\alpha_i,\psi^{(m)})f(\alpha_i|\psi^{(m)})d\alpha_i}
{\int f(Y_i,\tilde{T}_i,\Delta_i|\alpha_i,\psi^{(m)})f(\alpha_i|\psi^{(m)})d\alpha_i}.
\nonumber
\end{eqnarray}
The above integration can be computed via Gaussian quadrature\index{Gaussian quadrature} which approximates integrals by a weighted sum of target functions evaluated at prespecified sample points \citep{Vetterling:1989}.

In the M-step, $\psi$ is updated using
\begin{eqnarray}
\psi^{(m+1)}=argmax_{\Psi} ~ Q(\psi;\psi^{(m)}), \nonumber
\end{eqnarray}
where $Q(\psi; \psi^{(m)})=E_{\alpha|Y, \tilde{T}, \Delta, \psi^{(m)}}(\log L(\psi; Y, \tilde{T}, \Delta, \alpha))$. More details are provided in Supplementary material.

The proposed joint model is semiparametric because the baseline hazard function $\lambda(t)$ in Equation (\ref{surv}) is completely unspecified. This causes difficulty in estimating the standard errors of $\hat{\phi}$ and further drawing statistical inference. As a result, we employ the nonparametric bootstrap strategy to compute the variance of parameter estimates. Suppose $B$ bootstrap samples are formed by randomly sampling study subjects with replacement. The variance of $\hat{\phi}$ can be estimated using $1/(B - 1)\sum_{i = 1}^B (\phi^{(i)} - \bar{\phi})^2$, where $\phi^{(i)}$ is the estimate from the $i$-th bootstrap sample and $\bar{\phi}$ is the mean of $\phi^{(i)}$ over all $B$ samples.

\vspace{0.2cm}
Under conditions (C1)-(C5) (given in Supplementary material) and assuming B-spline knots are fixed, the following theorems establish the consistency and asymptotic normality for all the estimators in the joint model (\ref{long1})-(\ref{surv}). The results are derived using modern empirical process theory, and detailed proofs are deferred to Supplementary material.

\begin{theorem}
Under assumptions (C1)-(C5), the maximum likelihood estimator $(\phi,\Lambda)$ is strongly consistent under the product metric of the Euclidean norm and the supremum norm on $[0,\tau]$; that is,
\begin{equation*}
\|\hat \phi-\phi_0\|+\sup_{t\in[0,\tau]}|\hat\Lambda(t)-\Lambda_0(t)|\rightarrow 0~~~~~a.s.
\end{equation*}
\end{theorem}

\begin{theorem}
Under assumptions (C1)-(C5), $\sqrt{n}(\hat \phi-\phi_0,\hat\Lambda(t)-\Lambda_0(t))$ weakly converges to a Gaussian random element in $R^d\times l^{\infty}[0,\tau]$, where $d$ is the dimension of $\phi$ and $l^{\infty}[0,\tau]$ is the metric space of all bounded functions in $[0,\tau]$.
\end{theorem}

\subsection{Dynamic Prediction} \label{predicterror}

Since our joint model characterizes the association between biomarkers and event times, it can be used as a prognostic tool for dynamic prediction of event probabilities. Such prediction is conditional on biomarker measurements up to a given time point at which the event is yet to occur. Based on the predicted probability, physicians can better understand disease progression and make early decisions. We adopt the dynamic prediction accuracy measure developed by \citet{Proust:2009b} for latent-class joint models. This measure contrasts the predicted probability with observed data and thus can serve as an assessment tool to compare joint models for different sets of biomarkers in terms of their prediction accuracy.

For a new patient $i$ who has not experienced the event at time $s$, the objective is to predict the probability that $T_i \leq s + t$ conditional on $Y^{(s)}_i = \{Y_{ij}(t_{ijk}), t_{ijk} \leq s, j = 1, \ldots, J\}$ that contains all biomarker measurements up to time $s$. This probability is defined as
\begin{eqnarray}
P_i(s+t,s;\psi) &=& P(T_i\leq s+t|T_i > s,Y_i^{(s)};\psi). \label{dpred1}
\end{eqnarray}
Assuming $T_i \perp Y_i$ given $\alpha_i$, the probability (\ref{dpred1}) can be calculated as
\begin{eqnarray}
&&P(T_i\leq s+t|T_i > s,Y_i^{(s)};\psi) \nonumber \\
&& = \int P(T_i\leq s+t|T_i > s,Y_i^{(s)},\alpha_i;\psi)f(\alpha_i|T_i > s,Y_i^{(s)};\psi)d\alpha_i \nonumber \\
&& = \int P(T_i\leq s+t|T_i > s,\alpha_i;\psi)f(\alpha_i|T_i > s,Y_i^{(s)};\psi)d\alpha_i \nonumber \\
&& = \int \frac{S(s|\alpha_i;\psi)-S(s+t|\alpha_i;\psi)}{S(s|\alpha_i;\psi)}f(\alpha_i|T_i > s,Y_i^{(s)};\psi)d\alpha_i, \nonumber
\end{eqnarray}
where $S(\cdot|\alpha_i;\psi)$ is the survival function of $T_i$ conditional on $\alpha_i$. The posterior density $f(\alpha_i|T_i > s,Y_i^{(s)};\psi)$ is given by

\begin{eqnarray}
f(\alpha_i|T_i > s,Y_i^{(s)};\psi)&=&\frac{f(T_i>s,Y_i^{(s)},\alpha_i;\psi)}{f(T_i>s,Y_i^{(s)};\psi)} \nonumber \\
&=& \frac{S(s|\alpha_i;\psi)f(Y_i^{(s)}|\alpha_i;\psi)f(\alpha_i;\psi)}{\int S(s|\alpha_i;\psi)f(Y_i^{(s)}|\alpha_i;\psi)f(\alpha_i|\psi)d\alpha_i}. \nonumber
\end{eqnarray}

Note that $P_i(s + t, s; \psi)$ is calculated using $\hat{\psi}$ estimated from the joint model. Denote the dynamic prediction rule $\hat{S}(s + t|T_i > s, Y_i^{(s)}) = S(s + t|T_i > s, Y_i^{(s)}; \hat{\psi}) = 1 - P_i(s + t, s; \hat{\psi})$. The measure of predictive accuracy evaluates the prediction error at time $s + t$ conditional on the data history up to time $s$:
\begin{eqnarray}
&& \hat{err}_{Y,s}(s+t) \nonumber \\
&=& \frac{1}{N_s}\sum_{i=1}^{N_s}I(T_i>s+t)|1-\hat{S}(s + t|T_i > s, Y_i^{(s)})| \nonumber \\
&& + \Delta_iI(T_i\leq s+t)|0-\hat{S}(s + t|T_i > s, Y_i^{(s)})|\nonumber \\
&&  +(1-\Delta_i)I(T_i\leq s+t) [|1-\hat{S}(s + t|T_i > s, Y_i^{(s)})|\frac{\hat{S}(s + t|T_i > s, Y_i^{(s)})}{\hat{S}(T_i|T_i > s, Y_i^{(s)})} \nonumber \\
&& +|0-\hat{S}(s + t|T_i > s, Y_i^{(s)})|(1-\frac{\hat{S}(s + t|T_i > s, Y_i^{(s)})}{\hat{S}(T_i|T_i > s, Y_i^{(s)})})], \nonumber
\end{eqnarray}
where $N_s$ is the number of patients in the risk set at time $s$.

As stated previously, this measure of prediction accuracy can be used to compare across joint models with different or nested sets of covariates. Presumably, for pre-specified $s$ and $t$, including important biomarkers that are associated with the event of interest would reduce $\hat{err}_{Y,s}(s+t)$. Its use is illustrated in Section \ref{example}.

\section{REAL DATA ILLUSTRATION ON THE LUNG TRANSPLANT STUDY} \label{example}

This study consists of 215 patients who received lung transplantation at the University of California, Los Angeles between January 1, 2000 and December 31, 2010 \citep{Shino:2013}. Patients had a median follow-up of 3.5 years post-transplant. The primary endpoint of this study was Chronic Lung Allograft Dysfunction (CLAD), a major factor limiting long-term survival in lung transplant patients. CLAD was defined as a sustained drop of at least 20\% in the FEV${_1}$ from the average of the two best post-transplant FEV${_1}$ measurements. The study also recorded concentrations of CXCR3 chemokine ligands MIG and IP10 in bronchoalveolar lavage fluid post-transplant, with a range of 1 - 8 and a median of 3 observations per patient. MIG and IP10 are highly correlated ($r$ = 0.6) since they are in the same biological pathway. Our preliminary analyses showed that elevated MIG and IP10 concentrations were associated with an increased risk of CLAD in univariate Cox regression, but lost statistical significance when included simultaneously due to the fact that the strong correlation between MIG and IP10 resulted in unstable effect estimates. Also as stated previously, MIG and IP10 were intermittently measured and subject to substantial biological variation, which could introduce bias in Cox regression if they were treated as time-dependent covariates. To solve these issues, we use the model proposed in Section \ref{jointmodel} to assess predictive value of MIG and IP10 for the risk of CLAD. There were 108 (50.2\%) CLAD events and 611 MIG/IP10 observations. The median time from transplantation to the biomarker sampling date was 4.1 months.

We used cubic B-splines with evenly-spaced knots to estimate the mean and principal component functions in model (\ref{long2}) when fitting the joint model for CLAD and log$_2$-transformed MIG and IP10. Longitudinal measurements of MIG and IP10 were fit on the log-transformed time scale given that the measurements became more sparse as time progressed. There was a high biological variation in MIG and IP10, and no baseline patient demographic or clinical factors were significantly associated with these biomarkers. For illustrative purposes, we included patient age (in years) at the time of receiving transplant as a covariate in model (\ref{long1}). Baseline covariates age, male gender (male, yes/no), single lung transplant (single, yes/no), and idiopathic pulmonary fibrosis (ipf, yes/no) were included in the survival sub-model (\ref{surv}). We first considered $k = 2$ principal component functions in the latent process, and repeatedly fit the joint model with varying number of knots. Log-likelihood and AIC values for knots 2, 4, up to 12 are given in Table \ref{aictable}. The lowest AIC is achieved when there were 8 knots. The log-likelihood increased to -3030.4 when three principal component functions were considered, but the third principal component explained only 10\% of the total variation. This percentage was calculated using the estimated $D_{\kappa,\kappa}, \kappa = 1, 2, 3$, as discussed in Section \ref{modelsetup}. We are aware that it is a rough approximation of relative data variation given the fact that biomarker observation times varied from individual to individual, but a majority of the observations were obtained at about 1, 3, 6, and 12 months post-transplant. As a result, we choose the model with 8 knots and two principal component functions as our final model.

\begin{table}
\begin{center}
\def~{\hphantom{0}}
\caption{Log-likelihood and AIC for joint models with varying number of knots.}
{\begin{tabular}{cccc}
\hline \hline \\[-5pt]
Number of knots &Loglikelihood &Number of parameters &AIC \\
\hline \\[-5pt]
2 &-3208.31 &24 &6464.62 \\
4 &-3204.51 &30 &6469.03 \\
6 &-3192.87 &36 &6457.75 \\
8 &-3175.63 &42 &6435.26 \\
10 &-3170.04 &48 &6436.08 \\
12 &-3174.65 &54 &6457.30 \\
\hline \hline
\end{tabular}}
\label{aictable}
\end{center}
\end{table}

Maximum likelihood estimates of the final model are provided in Table \ref{jointfit} where the 95\% confidence intervals were calculated based on standard errors derived from 100 bootstrap samples and the p-values were calculated using the Wald's test. Since the estimated $\sigma_1^2$, $\sigma_2^2$, $D_{11}$, and $D_{22}$ from the bootstrap samples were right-skewed, their 95\% confidence intervals were calculated based on log-transformed estimates. Note that the two biomarkers MIG and IP10 were labeled as $j = 1, 2$, respectively. As expected, age is not associated with either MIG or IP10. The factor loading of $\mu_i(t)$ for IP10 ($\beta_{12}$) is 0.83 with a p-value of $<$ 0.0001, indicating a strong association between the two biomarkers. Conditional on the latent process $\mu(t)$, MIG has a little higher variability than IP10 as suggested by the estimates of $\sigma_1^2$ and $\sigma_2^2$. The first principal component function explains approximately 65\% of total variation in the latent process. The model detects a statistically significant association between the latent biomarker process and CLAD ($\gamma$ = 0.22, 95\% CI: 0.03 - 0.42, p = 0.0251); a higher level of MIG/IP10 is associated with an increased risk of CLAD, with a hazard ratio of 1.25 (95\% CI: 1.03 - 1.52). None of the demographic or clinical factors are associated with CLAD after controlling for the latent biomarker process. The estimated cumulative baseline hazard function (Figure \ref{chplot}) indicates that there is an approximately constant hazard of CLAD following lung transplantation. To fit this joint model, it took 23 minutes to obtain the point estimates of the parameters and 71 hours to complete 100 bootstrap.

The model estimated mean curve of MIG and IP10 is quite consistent with the empirical mean by LOESS method as shown in Figure \ref{rawandfitted}(a), though the fit tends to be poor at the lower and upper ends of data range. It is apparent that the IP10 concentration on average is lower than MIG, which explains why the factor loading $\beta_{12}$ of the latent process for IP10 is less than 1. Figure \ref{rawandfitted}(b) shows the estimated principal component functions $f_{\kappa}(t)$, $\kappa = 1, 2$. The first principal component function changes very little over time, acting similar to a random-intercept in linear mixed effect models. The second principal component varies around zero with no clear overall trend going up or down as time progresses, which is consistent with the fluctuations shown by the LOESS curves in Figure \ref{rawandfitted}(a). This principal component captures local biological variations in MIG and IP10 that are not accounted for by the first principal component. We also examined the distribution of estimated $\alpha_i$ from the joint model (Supplementary Figure 1). There is no obvious deviation from the normality assumption.

\begin{table}
\begin{center}
\def~{\hphantom{0}}
\caption{Parameter estimates from the joint model with 8 knots and 2 principal component functions. }
{\begin{tabular}{lrcr}
\hline \hline \\[-5pt]
Model parameter &Estimate &95\% CI &p-value \\
\hline \\[-5pt]
$\beta_{01}$ (MIG: age) &-0.02 &(-0.04, 0.01) &0.16 \\
$\beta_{02}$ (IP10: age) &-0.01 &(-0.03, 0.01) &0.15 \\
$\beta_{12}$ &0.83 &(0.76, 0.91) &$<$0.0001 \\
$\sigma_{1}^2$ &4.47 &(3.44, 5.18) & \\
$\sigma_{2}^2$ &2.87 &(2.15, 3.36) & \\
$D_{11}$ &28.5 &(6.5, 262.2) &  \\
$D_{22}$ &15.4 &(6.2, 29.6) &  \\
$\eta_{1}$ (age) &0.003 &(-0.02, 0.03) &0.80 \\
$\eta_{2}$ (male) &0.09 &(-0.34, 0.51) &0.69 \\
$\eta_{3}$ (single) &0.23 &(-0.29, 0.75) &0.39 \\
$\eta_{4}$ (ipf) &-0.41 &(-0.91, 0.10) &0.11 \\
$\gamma$ &0.22 &(0.03, 0.42) &0.0251 \\
\hline \hline
\end{tabular}}
\label{jointfit}
\end{center}
\end{table}

\begin{figure}
\begin{center}
\includegraphics[scale=0.53]{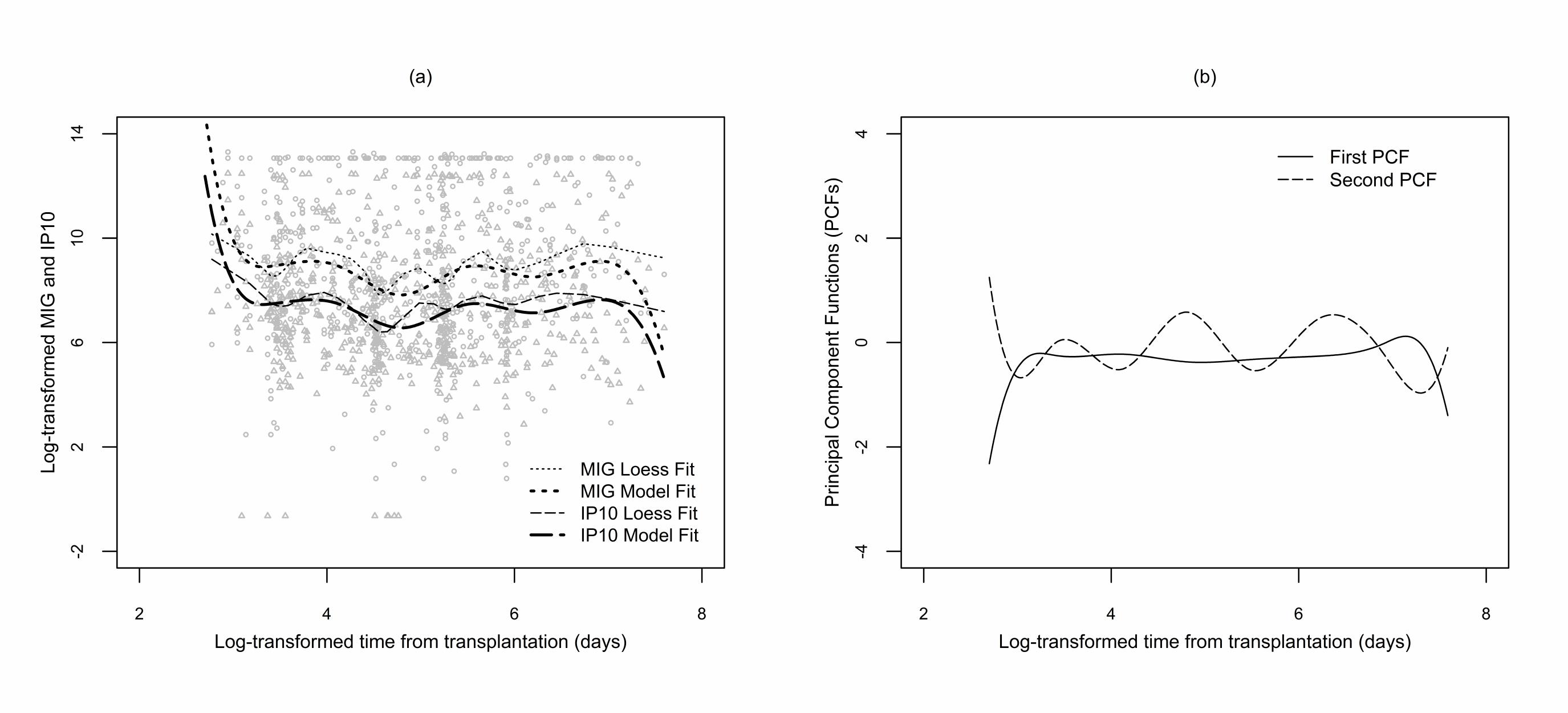}
\caption{(a) Joint model estimated mean curve (thick line) as compared to the LOESS curve (thin line). Observed data points for MIG and IP are labeled as circle and triangle, respectively. (b) Joint model estimated principal component functions (PCFs).}
\label{rawandfitted}
\end{center}
\end{figure}

\begin{figure}
\begin{center}
\includegraphics[scale=0.3]{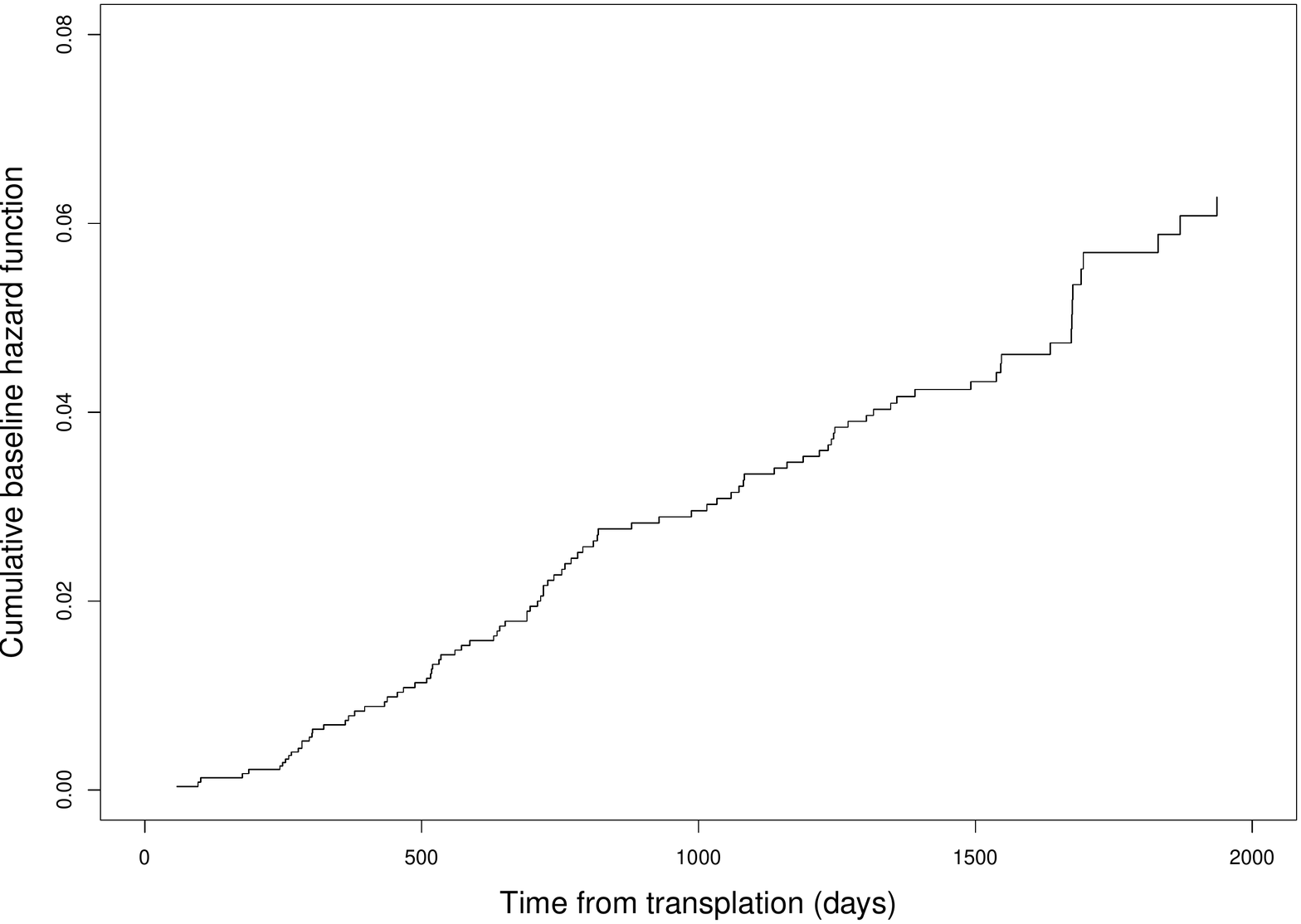}
\caption{Estimated cumulative baseline hazard function from the lung transplant study.}
\label{chplot}
\end{center}
\end{figure}

To assess the predictive accuracy of this joint model, $\hat{err}$ was calculated as given in Section \ref{predicterror}, using MIG and IP10 data up to 6 months post-transplant to predict CLAD occurrence in the next 1.5 years (i.e., in 2 years post-transplant), and was further compared with its counterparts in two other joint models using either MIG or IP10 alone to predict CLAD. These specific time points are clinically meaningful since the lung transplant patients are more closely monitored with the first half year following transplantation. The predictive accuracy measure $\hat{err}$ is 0.26 when MIG and IP10 were jointly modeled, and is 0.33 for both MIG and IP10 when they were modeled separately. This suggests that MIG and IP10 combined provide more accurate prediction for CLAD, and an improvement of 21\% is observed in this study.

\section{SIMULATION STUDIES}

We conducted a series of simulations to evaluate the performance of the proposed model and its estimation procedure under different settings, for which the results are summarized in Table \ref{various_models}. Specifically, we examined the impact of the level of variability or noise in the latent process (i.e., the variance of the random effects $\alpha_i$) on parameter estimation and also investigated whether the estimators were sensitive to mis-specification of the distributional assumption for the random effects. The following three scenarios were considered:
\begin{itemize}
\item Model 1 (low-medium variability, Gaussian random effects): The two dimensional random effects $\alpha_i$ associated with the latent process were simulated from Gaussian distributions with a variance of 2.0 and 1.0 for the first and second principal component functions, respectively. Subjects were followed on $t \in [0, 9]$, and two longitudinal outcomes were measured at every 0.5 time interval from baseline until the occurrence of the event or the end of the study. The baseline hazard was constant and set to value of 0.08 for all $t$, and the overall event rate was 78\%. Since the longitudinal outcomes were censored by the survival endpoint, the median number of visits per subject was 4. The knots of the B-splines for the latent process $\mu_i(t)$ were evenly spaced on $[0, 9]$ and the number of knots was set to 8, same as what was chosen for the lung transplant study discussed in Section \ref{example}. The covariate $X_i \sim N(0, 2)$ and was time-fixed.

\item Model 2 (high variability, Gaussian random effects): Similar to Model 1 with the following modifications. The random effects $\alpha_i$ were simulated from Gaussian distributions with a variance of 28.0 and 15.0, respectively. These values were chosen based on the estimates for $D_{11}$ and $D_{22}$ in the lung transplant study. In addition, the true values for the remaining parameters were also set close to the estimates in the lung transplant study to mimic the real data. The median number of visits per patient was 4.

\item Model 3 (medium-high variability, $Gamma$ random effects): Similar to Model 1 with the following modifications. The random effects $\alpha_i$ were simulated from $Gamma$ distributions with a variance of 12.0 and 8.0, respectively. This simulation aims to investigate whether the proposed estimation procedure is robust when the normality assumption for $\alpha_i$ is violated. Specifically, $\alpha_{i1}$ and $\alpha_{i2}$ were generated from $Gamma$ distributions with the shape and scale parameters set to (3,2) and (2,2), respectively. Both distributions are asymmetric with a longer right tail. Note that the variance of these random effects is at a level in between its counterpart in Models 1 and 2. We centered the random effects at the mean of their Gamma distributions to make sure the expectation was zero.
\end{itemize}

For these models we looked at a sample size of 215, same as that of the lung transplant study. A joint model with 8 knots was fit to each simulated dataset. We report the bias and standard deviation (SD) which was estimated using the empirical standard deviation of the estimates from 100 simulations. To illustrate how the estimators behave as the sample size increases, we performed simulations under the scenario of Model 1 with $n$  = 500, and the results are reported in Supplementary Table 1.

For the data generated from Model 1, the proposed estimation approach produces quite small bias for all the parameters except $D_{11}$, the variance of the random effect associated with the first principal component function, but this bias shrinks when the sample size increases to 500 (Supplementary Table 1). Figure \ref{chsimplot} indicates that the estimated cumulative baseline hazard functions from 20 randomly selected simulated datasets on average agree well with the true cumulative hazard.

Regarding the data that mimic the lung transplant study with a high level of variability in the latent proces (Model 2), we observe relatively small estimation bias and variability for the fixed effects $\beta_{01}$, $\beta_{02}$, and $\beta_{12}$ in the longitudinal sub-model and the coefficient $\gamma$ for the latent process in the survival sub-model, but the estimates for $D_{11}$ and $D_{22}$ show a larger bias and variation. In other words, although the high variability or noise in the latent process affects the estimation accuracy on $D_{11}$ and $D_{22}$, it does not seem to have a noticeable effect on the estimation of the fixed effects parameters $\beta_{01}$, $\beta_{02}$,  $\beta_{12}$ and $\gamma$. We also note that when the normality assumption for the random effects is violated (Model 3), the bias for $\beta$, $\gamma$, and the variance of the measurement error terms $\sigma_1^2$ and $\sigma_2^2$ remains small, suggesting that these parameter estimates seem to be robust to violation of the random effects normality assumption. Furthermore, the bias for $D_{11}$ and $D_{22}$ in Model 3 tends to be smaller than that in Model 2, suggesting again that the estimation accuracy for $D_{11}$ and $D_{22}$ would improve as the amount of noise in the latent process decreases.

We further investigated the sensitivity of parameter estimation with regards to the chosen knot number for the latent process. A series of joint models were fit to the data generated from Model 2 (an 8-knot model), but with the knot number set to 4, 6, 8, 10, and 12. As shown in Supplementary Table 2, the parameters $\beta_{01}$, $\beta_{02}$, $\beta_{12}$, and $\gamma$ consistently have small bias and variability regardless of the knot number specified. A larger bias is observed for $\sigma_1^2$ and $\sigma_2^2$ when the knot number is far below the true value, but this bias reduces as the knot number increases. The knot number has a marked impact on $D_{11}$ and $D_{22}$; both have a considerable bias and estimation variation when the knot number is either too small or large compared to the true value 8.

\begin{figure}
\begin{center}
\includegraphics[scale=0.4]{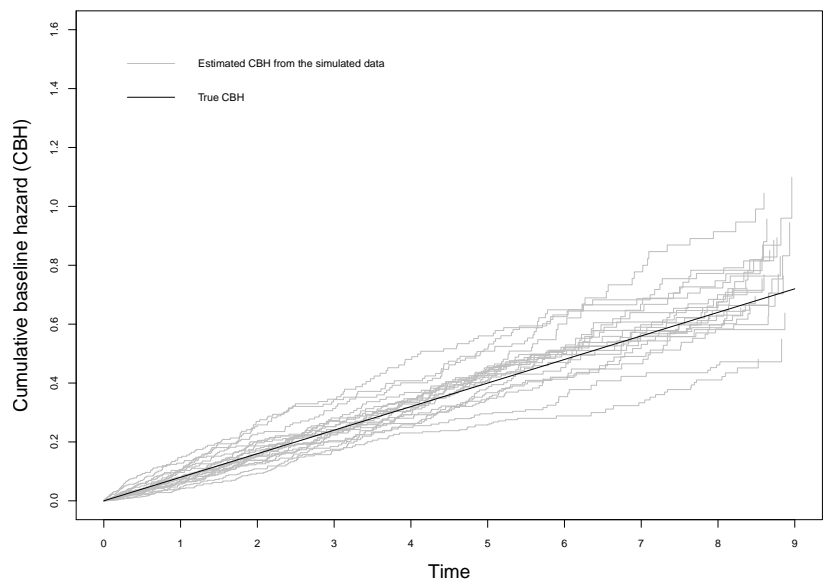}
\caption{Estimated cumulative baseline hazard functions from 20 simulated datasets (Model 1) as compared to the true cumulative baseline hazard.}
\label{chsimplot}
\end{center}
\end{figure}

\begin{table}\footnotesize
\begin{center}
\def~{\hphantom{0}}
\caption{Simulated estimation bias (and empirical standard deviation (SD)) for the joint model (\ref{long1})-(\ref{surv}) under three different scenarios. Each entry is based on 100 replications and the sample size is 215.}
{\begin{tabular}{ccrrrrrrrrr} \hline
 \\
&Parameter & $\beta_{01}$ &$\beta_{02}$ &$\beta_{12}$
&$\sigma_1^2$ &$\sigma_2^2$ &$D_{11}$ &$D_{22}$ &$\gamma$ \\
\hline \\[5pt]
Model 1 &\textit{True value} & \textit{-0.4} & \textit{-0.3} & \textit{0.8} & \textit{1.0} & \textit{1.0} & \textit{2.0}
&\textit{1.0} & \textit{0.5} \\[5pt]
(Low-medium variability, &Bias &0.003 &0.003 &$<$0.001 &-0.021 &-0.005 &0.189 &-0.033 &-0.013 \\
Gaussian random effects) &(SD) &(0.02) &(0.02) &(0.01) &(0.04) &(0.04) &(0.32) &(0.24) &(0.08) \\[10pt]

Model 2 &\textit{True value} & \textit{-0.02} & \textit{-0.01} & \textit{0.8} & \textit{4.5} & \textit{3.0}
& \textit{28.0} & \textit{15.0} & \textit{0.3} \\[5pt]
(High variability, &Bias &$<$0.001 &0.002 &\textcolor{blue}{-}0.014 &0.35 &0.21 &7.59 &-4.49 &-0.030 \\
Gaussian random effects)  &(SD) &($<$0.01) &($<$0.01) &(0.02) &(0.26) &(0.15) &(6.43) &(2.77) &(0.03) \\[10pt]

Model 3& \textit{True value} & \textit{-0.02} & \textit{-0.01} & \textit{0.8} & \textit{4.5} & \textit{3.0}
& \textit{12.0} & \textit{8.0} & \textit{0.3} \\[5pt]
(Medium-high variability, &Bias  &-0.001 &$<$0.001 &-0.007 &0.04 &0.03 &2.42 &-1.62 &-0.025  \\
$Gamma$ random effects)  &(SD)  &($<$0.01) &($<$0.01) &(0.01) &(0.20) &(0.12) &(2.31) &(1.68) &(0.05)  \\[5pt]
\hline \\
\end{tabular}}
\label{various_models}
\end{center}
\end{table}

\section{DISCUSSION}

We have developed a new flexible joint model for dynamic prediction of a clinically meaningful event from highly correlated biomarkers. We model the multiple biomarkers for each individual using a shared latent reduced rank longitudinal principal component model and then correlates the latent process to the event time of interest. B-splines are used to estimate the overall mean and principal component functions of the latent process. Our approach uses data to determine the functional form of the individual latent trajectories and thus is more flexible than the parametric latent trajectories used in previous work. We further establish asymptotic properties for semi-parametric maximum likelihood estimators and develop a
dynamic prediction tool under this model. As illustrated in the real data example, combining information from multiple highly correlated  biomarkers can lead to improved prediction accuracy for the event of interest than using a single biomarker alone. Our simulation results indicate the random effects variance parameters are often more difficult to estimate especially when there is substantial variation in the latent process, but they do not seem to have a noticeable effect on the estimation accuracy of the fixed effects $\beta$ and the association parameter $\gamma$ between the longitudinal and survival endpoints. In addition, estimation of the fixed effects in the longitudinal sub-model and the association between the latent process effect and the event risk is not very sensitive to mis-specification of the number of knots for the B-splines, whereas reasonable estimates of the principal component random effects may only be obtained when the assumed knot number is not far off from the value that can adequately capture data variation across time.

We note from  simulations that the parameter estimation seems to be quite robust to misspecification of random effects distributions, which is consistent with similar findings in the literature.   \citet{Song:2002} relaxed the normality assumption for the random effects in joint models and only required that the random effects have a smooth density. Their findings indicate that estimation of certain model parameters, including the fixed effects at the longitudinal endpoint, are remarkably robust to misspecification of the random effects distributions. Actually this robustness property may stem from linear mixed effects models as shown in a recent study by \citet{Schielzeth:2020} and related references therein. However, the theoretical basis of this phenomenon is still unclear, which warrants further research.

We point out that our approach and results have some limitations. First,  the estimator's asymptotic properties are derived by assuming the number of knots is fixed for the B-splines, whereas a full data-driven approach would ideally allow the flexibility of estimating the number of knots together with the remaining model parameters. However, proving the asymptotic properties under the latter scenario is challenging, which definitely warrants future research. Second, the maximum likelihood estimates of the model parameters are computed using an EM algorithm, in which the E-step involves intractable integrals that are approximated by Gaussian quadratures. The time cost of fitting such a model increases exponentially with the dimension of random effects, which is a well known computational challenge for most joint models. It would be of interest to develop more efficient computational methods, such as Laplace approximation \citep{Williamson:2018} or adaptive quadratures \citep{Tseng:2016}, to reduce the computational cost. Lastly, estimation of the parameters in the survival sub-model involves risk set assessment at all the observed event times, which leads to a computation complexity of $O(n^2)$ and is thus not scalable to super large $n$ data. Our group is currently engaged in developing more efficient algorithms for joint models of longitudinal and event time data and the findings will be reported in a sequel paper.

A further question is how to extend the current model to a more general setting where the biomarkers can be grouped into multiple biological pathways so that different pathways  have distinct trajectories, but biomarkers from the same pathway share similar trends over time.  Another related question is how to identify pathways and/or biomarkers that are most predictive of the clinical event. Further methodological developments are needed to answer these research questions. }

\begin{acknowledgement}
Gang Li's research was supported in part by National Institute of Health Grants
P30CA16042, UL1TR001881, and CA211015.
\end{acknowledgement}
\vspace*{1pc}

\noindent {\bf{Conflict of Interest}}

\noindent {\it{The authors have declared no conflict of interest.}}

%

\phantom{aaaa}
\end{document}


\DOIsuffix{bimj.200100000}
\Volume{52}
\Issue{61}
\Year{2010}
\pagespan{1}{}

\title[A Flexible Joint Model for  Multiple Longitudinal Biomarkers and A Time-to-Event Outcome]{A Flexible Joint Model for  Multiple Longitudinal Biomarkers and A Time-to-Event Outcome: With Applications to Dynamic Prediction Using Highly Correlated Biomarkers}
\author[Ning Li {\it{et al.}}]{Ning Li\inst{1}}
\address[\inst{1}]{Departments of Medicine and Biomathematics, University of California at Los Angeles, Los Angeles, California 90095, U.S.A.}
\author[]{Yi Liu\inst{2}}
\address[\inst{2}]{School of Mathematical Sciences, Ocean University of China, Qingdao 266100, China}
\author[]{Shanpeng Li\inst{3} }
\address[\inst{3}]{Department of Biostatistics, University of California at Los Angeles, Los Angeles, California 90095, U.S.A.}
\author[]{Robert M. Elashoff\inst{4} }
\address[\inst{4}]{Department of Biomathematics, University of California at Los Angeles, Los Angeles, California 90095, U.S.A.}
\author[]{Gang Li\footnote{Corresponding author: {\sf{e-mail: vli@ucla.edu}}, Phone: +00-310-206-5865, Fax: +00-310-267-2113}\inst{3} }
\Receiveddate{zzz} \Reviseddate{zzz} \Accepteddate{zzz}

\maketitle                   






\section*{Appendix}
\label{SM}
\renewcommand{\theequation}{A.\arabic{equation}}

Sections A.1 includes supplementary figures and tables, and Sections A.2 and A.3 include technical details for the EM algorithm and proofs of the theorems in Section 2.2.\\

\subsection*{A.1 Supplementary Figures and Tables}
\clearpage

\begin{figure}
\begin{center}
\includegraphics[scale=0.4]{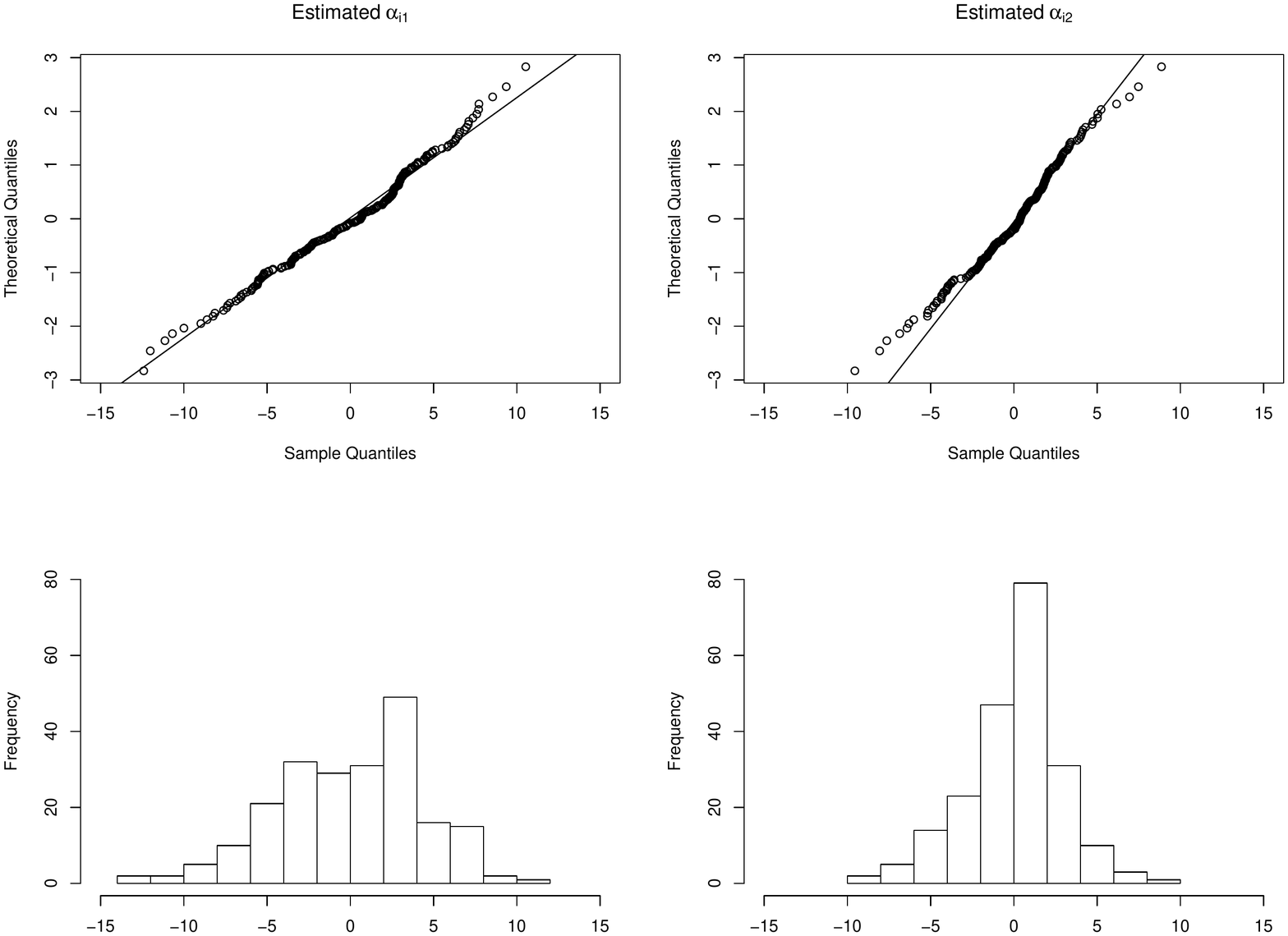}
\caption{Q-Q plot and histogram for estimated $\alpha_i$ for the lung transplant study.}
\label{qqplot}
\end{center}
\end{figure}

\clearpage

\begin{table}\footnotesize
\begin{center}
\def~{\hphantom{0}}
\caption{Simulated estimation bias (and empirical standard deviation (SD)) for Model 1 described in Section 4. Each entry is based on 100 replications (sample size = 500).}
{\begin{tabular}{crrrrrrrrr} \hline
 \\
Parameter & $\beta_{01}$ &$\beta_{02}$ &$\beta_{12}$
&$\sigma_1^2$ &$\sigma_2^2$ &$D_{11}$ &$D_{22}$ &$\gamma$ \\
\hline \\[5pt]
\textit{True value} & \textit{-0.4} & \textit{-0.3} & \textit{0.8} & \textit{1.0} & \textit{1.0} & \textit{2.0}
& \textit{1.0} & \textit{0.5} \\[5pt]
 &0.002 &$<$0.001 &$<$0.001 &-0.006 &$<$0.001 &0.116 &-0.063 &0.003 \\
 &(0.01) &(0.01) &(0.01) &(0.03) &(0.03) &(0.20) &(0.17) &(0.06) \\[5pt]
\hline \\
\end{tabular}}
\label{various_models}
\end{center}
\end{table}

\begin{table}\footnotesize
\begin{center}
\def~{\hphantom{0}}
\caption{Sensitivity analysis when a suboptimal number of knots is used: simulated estimation bias (and empirical standard deviation (SD)) for the joint model with 8 knots. Each entry is based on 100 replications (sample size = 215).}
{\begin{tabular}{crrrrrrrrr} \hline
 \\
Parameter & $\beta_{01}$ &$\beta_{02}$ &$\beta_{12}$
&$\sigma_1^2$ &$\sigma_2^2$ &$D_{11}$ &$D_{22}$ &$\gamma$ \\
(true value) & (-0.02) & (-0.01) & (0.8) & (4.5) & (3.0) & (28) & (15) & (0.3)\\
\hline \\
\# knots & & & & & & & & \\[5pt]
4 &0.003 &0.003 &-0.011 &1.25 &0.79 &32.94 &-10.95 &-0.059 \\
  &($<$0.01) &($<$0.01) &(0.02) &(0.34) &(0.21) &(72.78) &(3.46) &(0.03) \\[5pt]
6 &0.002 &0.003 &-0.014 &0.57 &0.34 &9.15 &-7.00 &-0.047 \\
  &($<$0.01) &($<$0.01) &(0.02) &(0.25) &(0.16) &(7.15) &(2.10) &(0.03) \\[5pt]
8 &$<$0.001 &0.002 &\textcolor{blue}{-}0.014 &0.35 &0.21 &7.59 &-4.49 &-0.030 \\
  &($<$0.01) &($<$0.01) &(0.02) &(0.26) &(0.15) &(6.43) &(2.77) &(0.03) \\[5pt]
10 &0.002 &0.003 &-0.018 &0.38 &0.23 &9.15 &-3.25 &-0.036 \\
   &($<$0.01) &($<$0.01) &(0.02) &(0.22) &(0.14) &(6.87) &(3.17) &(0.03) \\[5pt]
12 &0.004 &0.004 &-0.017 &0.36 &0.22 &6.53 &-2.47 &-0.032 \\
   &($<$0.01) &($<$0.01) &(0.02) &(0.31) &(0.22) &(8.24) &(3.53) &(0.03) \\[5pt]
\hline \\
\end{tabular}}
\label{tableknots}
\end{center}
\end{table}

\clearpage

\subsection*{A.2 The EM Algorithm}

E-step: in the E-step of the $(m+1)$-th iteration, conditional on the observed data and the parameter estimates from the $m$-th iteration, we evaluate
\begin{eqnarray*}
E\{h(\alpha_i)|Y_i,\tilde{T}_i,\Delta_i,\psi^{(m)}\}&=&\int h(\alpha_i)f(\alpha_i|Y_i,\tilde{T}_i,\Delta_i,\psi^{(m)})d\alpha_i \nonumber \\
&=&\frac{\int h(\alpha_i)f(\alpha_i,Y_i,\tilde{T}_i,\Delta_i|\psi^{(m)})d\alpha_i}{f(Y_i,\tilde{T}_i,\Delta_i|\psi^{(m)})} \nonumber \\
&=&\frac{\int h(\alpha_i)f(Y_i,\tilde{T}_i,\Delta_i|\alpha_i,\psi^{(m)})f(\alpha_i|\psi^{(m)})d\alpha_i}
{\int f(Y_i,\tilde{T}_i,\Delta_i|\alpha_i,\psi^{(m)})f(\alpha_i|\psi^{(m)})d\alpha_i}.
\end{eqnarray*}
The integrals can be evaluated using Gaussian-Hermite quadrature.

\noindent M-step: the complete-data log-likelihood is
\begin{eqnarray*}
l(\psi,Y_i,\tilde{T}_i,\Delta_i)&=&{\sum_{i=1}^n }\Big[\sum_{j=1}^J\sum_{u=1}^{n_i}\Big\{-\frac{1}{2}\log 2\pi-\frac{1}{2}\log \sigma_j^2-\frac{1}{2\sigma_j^2}(Y_{ij}(t_{iu})-X_{ij}(t_{iu})^\top \beta_{0j}\\
&&-b(t_{iu})^\top \theta\beta_{1j}-b(t_{iu})^\top \Theta\alpha_i\beta_{1j})^2\Big\}\\
&&+\Delta_i(\log\lambda_0(\tilde T_i)+Z_i(\tilde T_i)^\top \eta + b(\tilde T_i)^\top\theta\gamma+b(\tilde T_i)^\top\Theta\alpha_i\gamma)\\
&&-\int_0^{\tilde T_i}\lambda_0(t)\exp(Z_i(t)^\top \eta + b(t)^\top\theta\gamma+b(t)^\top\Theta\alpha_i\gamma)dt\Big]\\
&&+\sum_{i=1}^n\sum_{k=1}^{K}\Big[-\frac{1}{2}\log 2\pi-\frac{1}{2}\log D_{k,k}-\frac{1}{2D_{k,k}}\alpha_{ik}^2\Big].
\end{eqnarray*}

Use E to stand for $E\{\cdot|Y_i,\tilde{T}_i,\Delta_i,\psi^{(m)}\}$. We have
\begin{eqnarray*}
\sigma_j^{2(m+1)} &=& \frac{1}{\sum_{i=1}^n n_i}\sum_{i=1}^n\sum_{u=1}^{n_i}E(Y_{ij}(t_{iu})-X_{ij}(t_{iu})^\top \beta_{0j}^{(m)}-b(t_{iu})^\top \theta^{(m)}\beta_{1j}^{(m)}-b(t_{iu})^\top \Theta^{(m)}\alpha_i\beta_{1j}^{(m)})^2,\\
D_{k,k}^{(m+1)} &=& \frac{1}{n}\sum_{i=1}^nE(\alpha_{ik}^2),~~k=1,\cdots,K,\\
\beta_{1j}^{(m+1)} &=& \frac{\sum_{i=1}^n\sum_{u=1}^{n_i}E(Y_{ij}(t_{iu})-X_{ij}(t_{iu})^\top \beta_{0j}^{(m)})(b(t_{iu})^\top \theta^{(m)}+b(t_{iu})^\top \Theta^{(m)}\alpha_i)}{\sum_{i=1}^n\sum_{u=1}^{n_i}E(b(t_{iu})^\top \theta^{(m)}+b(t_{iu})^\top \Theta^{(m)}\alpha_i)^2},\\
&&j=2,\cdots,J.
\end{eqnarray*}
\\
Let $d(t_u)$ be the number of failures at time $t_u$, and $R(t_u)$ be the risk set at time $t_u$. The cumulative baseline hazard function is
\begin{eqnarray*}
\Lambda_0^{(m+1)}(t)=\sum_{t_u\leq t}\frac{d(t_u)}{\sum_{r\in R(t_u)}\exp(Z_r(t_u)^\top \eta^{(m)}+b(t_u)^\top\theta^{(m)}\gamma^{(m)})E\exp(b(t_u)^\top\Theta^{(m)}\alpha_i\gamma^{(m)})}.
\end{eqnarray*}
\\
No closed-form solutions exist for $\beta_{0j}$, $\eta$, $\gamma$, $\theta$ and $\Theta$, which are updated by a one-step Newton-Raphson algorithm in each iteration:
\begin{eqnarray*}
\beta_{0j}^{(m+1)}=\beta_{0j}^{(m)}+S_{\beta_{0j}}^{(m)}/I_{\beta_{0j}}^{(m)},
\end{eqnarray*}
where $j=1,\cdots,J$, with $S_{\beta_{0j}}^{(m)}$ and $I_{\beta_{0j}}^{(m)}$ being
\begin{eqnarray*}
S_{\beta_{0j}}^{(m)}&=&\sum_{i=1}^n\sum_{u=1}^{n_i}\frac{1}{\sigma_j^{2(m)}}E(Y_{ij}(t_{iu})-X_{ij}(t_{iu})^\top \beta_{0j}^{(m)}-b(t_{iu})^\top \theta^{(m)}\beta_{1j}^{(m)}\\
&&-b(t_{iu})^\top \Theta^{(m)}\alpha_i\beta_{1j}^{(m)})X_{ij}(t_{iu}), \nonumber
\end{eqnarray*}
\begin{eqnarray*}
I_{\beta_{0j}}^{(m)}&=&\sum_{i=1}^n\sum_{u=1}^{n_i}\frac{1}{\sigma_j^{2(m)}}X_{ij}(t_{iu})X_{ij}(t_{iu})^\top,
\nonumber
\end{eqnarray*}

\begin{eqnarray*}
\eta^{(m+1)}=\eta^{(m)}+S_{\eta}^{(m)}/I_{\eta}^{(m)},
\end{eqnarray*}
with $S_{\eta}^{(m)}$ and $I_{\eta}^{(m)}$ being
\begin{eqnarray*}
S_{\eta}^{(m)}&=&\sum_{i=1}^n\Big\{\Delta_iZ_i(\tilde T_i)-\sum_{t_u\leq \tilde T_i}\lambda_0(t_u)\exp(Z_i(t_u)^\top \eta^{(m)}\\
&&+b(t_u)^\top \theta^{(m)}\gamma^{(m)}) E\exp(b(t_u)^\top\Theta^{(m)}\alpha_i\gamma^{(m)})Z_i(t_u)\Big\}, \nonumber
\end{eqnarray*}
\begin{eqnarray*}
I_{\eta}^{(m)}&=&\sum_{i=1}^n\sum_{t_u\leq \tilde T_i}\lambda_0(t_u)\exp(Z_i(t_u)^\top \eta^{(m)}+b(t_u)^\top \theta^{(m)}\gamma^{(m)})\\
&&\times E\exp(b(t_u)^\top\Theta^{(m)}\alpha_i\gamma^{(m)})Z_i(t_u)Z_i(t_u)^\top, \nonumber
\end{eqnarray*}

\begin{eqnarray*}
\gamma^{(m+1)}=\gamma^{(m)}+S_{\gamma}^{(m)}/I_{\gamma}^{(m)},
\end{eqnarray*}
with $S_{\gamma}^{(m)}$ and $I_{\gamma}^{(m)}$ being
\begin{eqnarray*}
S_{\gamma}^{(m)}&=&\sum_{i=1}^n\Big\{\Delta_i(b(\tilde T_i)^\top \theta^{(m)}+Eb(\tilde T_i)^\top\Theta^{(m)}\alpha_i)\\
&&-\sum_{t_u\leq \tilde T_i}\lambda_0(t_u)\exp(Z_i(t_u)^\top \eta^{(m)}+b(t_u)^\top \theta^{(m)}\gamma^{(m)})\\
&&\times E\exp(b(t_u)^\top\Theta^{(m)}\alpha_i\gamma^{(m)})(b(t_u)^\top \theta^{(m)}+b(t_u)^\top\Theta^{(m)}\alpha_i)\Big\}, \nonumber
\end{eqnarray*}
\begin{eqnarray*}
I_{\gamma}^{(m)}&=&\sum_{i=1}^n\sum_{t_u\leq \tilde T_i}\lambda_0(t_u)\exp(Z_i(t_u)^\top \eta^{(m)}+b(t_u)^\top \theta^{(m)}\gamma^{(m)})\\
&&\times E\exp(b(t_u)^\top\Theta^{(m)}\alpha_i\gamma^{(m)})(b(t_u)^\top \theta^{(m)}+b(t_u)^\top\Theta^{(m)}\alpha_i)^2,
\nonumber
\end{eqnarray*}

\begin{eqnarray*}
\theta^{(m+1)}=\theta^{(m)}+S_{\theta}^{(m)}/I_{\theta}^{(m)},
\end{eqnarray*}
with $S_{\theta}^{(m)}$ and $I_{\theta}^{(m)}$ being
\begin{eqnarray*}
S_{\theta}^{(m)}&=&\sum_{i=1}^n\Big[\sum_{j=1}^J\sum_{u=1}^{n_i}\Big\{\frac{1}{\sigma_j^{2(m)}}(Y_{ij}(t_{iu})-X_{ij}(t_{iu})^\top \beta_{0j}^{(m)}\\
&&-b(t_{iu})^\top \theta^{(m)}\beta_{1j}^{(m)}-b(t_{iu})^\top \Theta^{(m)} E\alpha_i\beta_{1j}^{(m)})b(t_{iu})\beta_{1j}^{(m)}\Big\}\\
&&+\Delta_ib(\tilde T_i)\gamma^{(m)}-\sum_{t_u\leq \tilde T_i}\lambda_0(t_u)\exp(Z_i(t_u)^\top \eta^{(m)}\\
&&+b(t_u)^\top \theta^{(m)}\gamma^{(m)})E\exp(b(t_u)^\top\Theta^{(m)}\alpha_i\gamma^{(m)})b(t_u)\gamma^{(m)}\Big], \nonumber
\end{eqnarray*}
\begin{eqnarray*}
I_{\theta}^{(m)}&=&\sum_{i=1}^n\Big[\sum_{j=1}^J\sum_{u=1}^{n_i}\frac{1}{\sigma_j^{2(m)}}b(t_{iu})b(t_{iu})^\top \beta_{1j}^{2(m)}\\
&&+\sum_{t_u\leq \tilde T_i}\lambda_0(t_u)\exp(Z_i(t_u)^\top \eta^{(m)}+b(t_u)^\top \theta^{(m)}\gamma^{(m)})\\
&&\times E\exp(b(t_u)^\top\Theta^{(m)}\alpha_i\gamma^{(m)})b(t_u)b(t_u)^\top\gamma^{2(m)}\Big]. \nonumber
\end{eqnarray*}

We separate $\Theta=(\Theta_1,\cdots,\Theta_K)$, where $\Theta_w$, $w=1,\cdots,K$ is a $q\times 1$ vector:
\begin{eqnarray*}
\Theta_w^{(m+1)}=\Theta_w^{(m)}+S_{\Theta_w}^{(m)}/I_{\Theta_w}^{(m)},
\end{eqnarray*}
with $S_{\Theta_w}^{(m)}$ and $I_{\Theta_w}^{(m)}$ being
\begin{eqnarray*}
S_{\Theta_w}^{(m)}&=&\sum_{i=1}^n\Big[\sum_{j=1}^J\sum_{u=1}^{n_i}\Big\{\frac{1}{\sigma_j^2}E(Y_{ij}(t_{iu})-X_{ij}(t_{iu})^\top \beta_{0j}^{(m)}-b(t_{iu})^\top \theta^{(m)}\beta_{1j}^{(m)}\\
&&-\sum_{w=1}^Kb(t_{iu})^\top \Theta_w^{(m)} \alpha_{iw}\beta_{1j}^{(m)})b(t_{iu})\alpha_{iw}\beta_{1j}^{(m)}\Big\}+\Delta_ib(\tilde T_i)E\alpha_{iw}\gamma^{(m)}\\
&&-\sum_{t_u\leq \tilde T_i}\lambda_0(t_u)\exp(Z_i(t_u)^\top \eta^{(m)}+b(t_u)^\top \theta^{(m)}\gamma^{(m)})\\
&&\times E\exp(\sum_{w=1}^Kb(t_u)^\top\Theta_w^{(m)}\alpha_{iw}\gamma^{(m)})b(t_u)\alpha_{iw}\gamma^{(m)}\Big],
\nonumber
\end{eqnarray*}
\begin{eqnarray*}
I_{\Theta_w}^{(m)}&=&\sum_{i=1}^n\Big[\sum_{j=1}^J\sum_{u=1}^{n_i}\frac{1}{\sigma_j^{2(m)}}b(t_{iu})b(t_{iu})^\top E\alpha_{iw}^2\beta_{1j}^{2(m)}\\
&&+\sum_{t_u\leq \tilde T_i}\lambda_0(t_u)\exp(Z_i(t_u)^\top \eta^{(m)}+b(t_u)^\top \theta^{(m)}\gamma^{(m)})\\
&&\times E\exp(\sum_{w=1}^Kb(t_u)^\top\Theta_w^{(m)}\alpha_{iw}\gamma^{(m)})b(t_u)b(t_u)^\top\alpha_{iw}^{2(m)}\gamma^{2(m)}\Big].
\nonumber
\end{eqnarray*}

\clearpage
\subsection*{A.3 Proofs of Theorems}

The following assumptions are needed to prove the theorems:

(C1) Recall that $n_i$ is the number of observed repeated measurements. There exists an integer $n_0$ such that $P(n_i\leq n_0)=1$. Moreover, conditional on $T$, and the longitudinal covariate history until to time $T$, the probability of $n_i>k$ is greater than zero with probability one.

(C2) The maximal right-censoring time is equal to $\tau$.

(C3) For any $t\in[0,\tau]$, the covariate process $\mathcal{X}(t)$ is fully observed and conditional on $\alpha$, the longitudinal covariate history prior to time $t$, the longitudinal response history prior to time $t$, and $T\ge t$, the distribution of $\mathcal{X}(t)$ depends only on the longitudinal covariate history prior to time $t$. Moreover, with probability one, $\mathcal{X}(t)$ is continuously differentiable in $[0,\tau]$ and $\max_{t\in[0,\tau]}\|\mathcal{X}'(t)\|<\infty$, where $\|\cdot\|$ denotes the Euclidean norm in real space and $\mathcal{X}'(t)$ denotes the derivative of $\mathcal{X}(t)$ with respect to $t$.

(C4) $x\beta_0 + b(\beta_1\otimes\theta)=0$, then $\beta_0=0$, $\beta_1=0$ and $\theta=0$. Additionally, if there exist constant vectors $C_0$ and $\tilde C_0$ such that with positive probability, $Z(t)^\top C_0=\nu(t)$,  $b(t)^\top \tilde C_0=\tilde\nu(t)$ for deterministic functions $\nu(t)$ and $\tilde\nu(t)$ for all $t\in [0,\tau]$, then $C_0=0$, $\tilde C_0=0$, $\nu(t)=0$, and $\tilde\nu(t)=0$.

(C5) The true parameter for $\phi$, denoted by $\phi_0$ satisfies $\|\phi_0\|\leq M_0$ for a known positive constant $M_0$. Moreover, the true hazard rate function, denoted by $\lambda_0(t)$, is bounded and positive in $[0,\tau]$.

\ \
Below we introduce some further notations needed for the proof of the main theorem. The marker process $Y(t)$ satisfies
\begin{eqnarray*}
Y_{ij}(t)=X_{ij}^\top(t)\beta_{0j}+b(t)^\top\theta\beta_{1j}+b(t)^\top\Theta\alpha_i\beta_{1j}+\epsilon_{ij}(t),
\end{eqnarray*}
where each component of the model is defined in the manuscript. Suppose the observed times of each patient are $t_{i1}$, $t_{i2}$, $\cdots$, $t_{in_i}$. Define $y_{ij} = (Y_{ij}(t_{i1}),\cdots,Y_{ij}(t_{in_i}))^\top$ which is a $n_i\times 1$ vector, $x_{ij}=(X_{ij}(t_{i1}),\cdots,X_{ij}(t_{in_i}))^\top$ a $n_i\times p_j$ matrix, $\tilde b_i = (b(t_{i1}),\cdots,b(t_{in_i}))^\top$ a $n_i\times q$ matrix, and $\epsilon_{ij}=(\epsilon_{ij}(t_{i1}),\cdots,\epsilon_{ij}(t_{in_i}))^\top$ a $n_i\times 1$ vector. The above model can be rewritten as
\begin{eqnarray*}
y_{ij}=x_{ij}\beta_{0j}+b_i\theta\beta_{1j}+b_i\Theta\alpha_i\beta_{1j}+\epsilon_{ij}.
\end{eqnarray*}
Further define $y_i=(y_{i1}^\top,\cdots,y_{iJ}^\top)^\top$ which is a $n_iJ\times 1$ vector, $x_i = diag(x_{i1},\cdots,x_{iJ})$ a $n_iJ\times \sum_{j=1}^J p_j$ matrix, $b_i = diag(\tilde b_{i},\cdots,\tilde b_{i})$ a $n_iJ\times qJ$ matrix, $\beta_0 = (\beta_{01}^\top,\cdots,\beta_{0J}^\top)^\top$ a $\sum_{j=1}^J p_j\times 1$ vector, $\beta_1 = (\beta_{11},\cdots,\beta_{1J})^\top$ a $J\times 1$ dimensional vector, and $\epsilon_i=(\epsilon_{i1}^\top,\cdots,\epsilon_{iJ}^\top)^\top$ a $n_iJ\times 1$ vector. The marker model can then be written as
\begin{eqnarray}
y_{i}=x_{i}\beta_{0}+b_i(\beta_1\otimes\theta)+b_i(\beta_1\otimes\Theta)\alpha_i+\epsilon_{i},
\end{eqnarray}
and the hazard function is
\begin{eqnarray}
\lambda(t)\exp(Z_i(t)^\top \eta+b(t)^\top \theta\gamma +b(t)^\top \Theta \alpha_i \gamma),
\end{eqnarray}
where the baseline function $\lambda(t)$ is unspecified.

The observed likelihood function is proportional to
\begin{eqnarray}\label{L}
&&L(\psi;Y,\tilde T,\Delta) \nonumber\\
&&= \Pi_{i=1}^n\int_{\alpha}\frac{1}{(2\pi)^{n_iJ/2}|\Sigma|^{1/2}}\nonumber\\
&&\times\exp(-\frac{1}{2}(y_{i}-x_{i}\beta_{0}-b_i(\beta_1\otimes\theta)-b_i(\beta_1\otimes\Theta)\alpha))^\top \Sigma^{-1}\nonumber\\
&&(y_{i}-x_{i}\beta_{0}-b_i(\beta_1\otimes\theta)-b_i(\beta_1\otimes\Theta)\alpha))\nonumber\\
&&\times\lambda(\tilde T_i)^{\Delta_i}\exp(\Delta_i(Z_i(\tilde T_i)^\top \eta+b(\tilde T_i)^\top \theta\gamma +b(\tilde T_i)^\top \Theta \alpha \gamma)\nonumber\\
&&-\int_{0}^{\tilde T_i}\exp(Z_i(t)^\top \eta+b(t)^\top \theta\gamma +b(t)^\top \Theta \alpha \gamma)d\Lambda(t))\nonumber\\
&&\times\frac{1}{(2\pi)^{k/2}|D|^{1/2}}\exp(-\frac{1}{2}\alpha^\top D^{-1}\alpha)d\alpha.
\end{eqnarray}

For convenience, we use $O$ to abbreviate the observed statistics. Let
\begin{eqnarray*}
&&G(\alpha,O,\phi,\Lambda)\\
&=&\frac{1}{(2\pi)^{n_iJ/2}|\Sigma|^{1/2}}\frac{1}{(2\pi)^{k/2}| D|^{1/2}}\\
&&\times\exp\Big(-\frac{1}{2}(y_{i}-x_{i}\beta_{0}-b_i(\beta_1\otimes\theta)-b_i(\beta_1\otimes\Theta)\alpha))^\top \\
&&\times\Sigma^{-1}(y_{i}-x_{i}\beta_{0}-b_i(\beta_1\otimes\theta)-b_i(\beta_1\otimes\Theta)\alpha)\\
&&-\frac{1}{2}\alpha^\top  D^{-1}\alpha+\Delta_i(Z_i(\tilde T_i)^\top \eta+b(\tilde T_i)^\top\theta\gamma +b(\tilde T_i)^\top \Theta \alpha \gamma)\\
&&-\int_{0}^{\tilde T_i}\exp(Z_i(t)^\top \eta+b(t)^\top \theta\gamma  +b(t)^\top \Theta \alpha \gamma)d\Lambda(t))\Big),
\end{eqnarray*}
and
\begin{eqnarray*}
&&Q(t,O,\phi,\Lambda)=\frac{\int_{\alpha}G(\alpha,O,\phi,\Lambda)\exp(Z_i(t)^\top \eta+b(t)^\top \theta\gamma +b(t)^\top \Theta \alpha \gamma)d\alpha}{\int_{\alpha}G(\alpha,O,\phi,\Lambda)d\alpha}.
\end{eqnarray*}
We further define a class $\mathcal{F}=\{Q(t,O,\phi,\Lambda):t\in[0,\tau],\phi\in\Phi,\Lambda\in \mathcal{Z},\Lambda(0)=0,\Lambda(\tau)\leq B_0\}$, where $\mathcal{Z}$ contains all nondecreasing functions in $[0,\tau]$.

\vspace{1cm}
\begin{lemma}
\label{lemma1}
Under assumptions (C1)-(C5), $\mathcal{F}$ is P-Donsker.
\end{lemma}

\textbf{Proof of Lemma \ref{lemma1}}. We can rewrite $Q(t,O,\phi,\Lambda)$ as
\begin{eqnarray*}
Q(t,O,\phi,\Lambda) = Q_1(t,O,\phi)\frac{Q_2(t,O,\phi,\Lambda)}{Q_3(t,O,\phi,\Lambda)},
\end{eqnarray*}
where
\begin{eqnarray*}
&&Q_1(t,O,\phi)\\
&&=\exp(Z(t)^\top \eta+b(t)^\top \theta\gamma+\frac{1}{2}\Theta^\top b(t) \gamma V^{-1}b(t)^\top\Theta \gamma\\
&&+b(t)^\top \Theta \gamma V^{-1}((b(\beta_1\otimes\Theta))^\top\Sigma^{-1}(y-x\beta_{0}-b(\beta_1\otimes\theta))+(\Delta b(\tilde T)^\top \Theta \gamma)^\top),
\end{eqnarray*}
\begin{eqnarray*}
&&Q_2(t,O,\phi,\Lambda)\\
&&=\int_{\alpha}\exp\Big(-\frac{1}{2}\alpha^\top\alpha-\int_0^{\tilde T} \exp(Z(t)^\top \eta+b(t)^\top \theta\gamma+b(t)^\top \Theta\gamma V^{-1/2}\alpha \\ &&+b(t)^\top \Theta\gamma V^{-1}((b(\beta_1\otimes\Theta))^\top\Sigma^{-1}(y-x\beta_{0}-b(\beta_1\otimes\theta))\\
&&+(\Delta b(\tilde T)^\top \Theta \gamma)^\top+(b(t)^\top \Theta \gamma)^\top)d\Lambda(t)\Big)d\alpha,
\end{eqnarray*}
and
\begin{eqnarray*}
&&Q_3(t,O,\phi,\Lambda)\\
&&=\int_{\alpha}\exp\Big(-\frac{1}{2}\alpha^\top\alpha-\int_0^{\tilde T}\exp(Z(t)^\top \eta+b(t)^\top \theta\gamma+b(t)^\top \Theta\gamma V^{-1/2}\alpha \\
&&~~+ b(t)^\top \Theta\gamma V^{-1}((b(\beta_1\otimes\Theta))^\top\Sigma^{-1}(y-x\beta_{0}-b(\beta_1\otimes\theta))\\
&&~~+(\Delta b(\tilde T)^\top \Theta \gamma)^\top)d\Lambda(t)\Big)d\alpha.
\end{eqnarray*}
We can easily show that $Q_1(t,O,\phi)$ is continuously differentiable with respect to $t$ and $\phi$ and
\begin{eqnarray*}
\|\nabla_{\phi}Q_1(t,O,\phi)\|+|\frac{d}{dt}Q_1(t,O,\phi)|\leq \exp(g_1+g_2\|y\|)
\end{eqnarray*}
for some constants $g_1$ and $g_2$. Furthermore, it holds that
\begin{eqnarray*}
&&\|\nabla_{\phi}Q_2(t,O,\phi,\Lambda)\|+|\frac{d}{dt}Q_2(t,O,\phi,\Lambda)|\\
&&\leq \int_{\alpha}\exp(-\frac{1}{2}\alpha^\top\alpha)e^{g_3\|\alpha\|+g_4\|y\|+g_5}B_0d\alpha\\
&&\leq \exp(g_6+g_7\|y\|)
\end{eqnarray*}
and $\|\nabla_{\phi}Q_3(t,O,\phi,\Lambda)\|+|\frac{d}{dt}Q_3(t,O,\phi,\Lambda)|\leq \exp(g_8+g_9\|y\|)$ for some positive constants $g_3,\cdots,g_9$. Additionally,
\begin{eqnarray*}
&&|Q_2(t,O,\phi,\Lambda_1)-Q_2(t,O,\phi,\Lambda_2)|\\
&\leq& \Big|\int_{\alpha}\exp(-\frac{1}{2}\alpha^\top\alpha)\int_0^{\tilde T} \exp(Z(t)^\top \eta+b(t)^\top \theta\gamma+b(t)^\top \Theta\gamma V^{-1/2}\alpha \\
&&~~~~+ b(t)^\top \Theta\gamma V^{-1}((b(\beta_1\otimes\Theta))^\top\Sigma^{-1}(y-x\beta_{0}-b(\beta_1\otimes\theta))\\
&&~~~~~~~~+(\Delta b(\tilde T)^\top \Theta \gamma)^\top+(b(t)^\top \Theta \gamma)^\top)d(\Lambda_1-\Lambda_2)(t)d\alpha\Big|\\
&\leq& (2\pi)^{k/2}\Big|\int_0^{\tilde T} \exp(Z(t)^\top \eta+b(t)^\top \theta\gamma+b(t)^\top \Theta\gamma V^{-1}\gamma\Theta^\top b(t)/2 \\
&&~~~~+ b(t)^\top \Theta\gamma V^{-1}((b(\beta_1\otimes\Theta))^\top\Sigma^{-1}(y-x\beta_{0}-b(\beta_1\otimes\theta))\\
&&~~~~~~~~+(\Delta b(\tilde T)^\top \Theta \gamma)^\top+(b(t)^\top \Theta \gamma)^\top)d(\Lambda_1-\Lambda_2)(t)\Big|\\
&\leq&(2\pi)^{k/2}|\Lambda_1(\tilde T)-\Lambda_2(\tilde T)|\exp(Z(t)^\top \eta+b(t)^\top \theta\gamma+b(t)^\top \Theta\gamma V^{-1}\gamma\Theta^\top b(t)/2 \\
&&~~~~+ b(t)^\top \Theta\gamma V^{-1}((b(\beta_1\otimes\Theta))^\top\Sigma^{-1}(y-x\beta_{0}-b(\beta_1\otimes\theta))+(\Delta b(\tilde T)^\top \Theta \gamma)^\top+(b(t)^\top \Theta \gamma)^\top)\\
&&~~+(2\pi)^{k/2}\int_0^{\tilde T}|\Lambda_1(t)-\Lambda_2(t)\Big|\frac{d}{dt}\exp(Z(t)^\top \eta+b(t)^\top \theta\gamma+b(t)^\top \Theta\gamma V^{-1}\gamma\Theta^\top b(t)/2 \\
&&~~~~~~~~~~~~~~~+ b(t)^\top \Theta\gamma V^{-1}((b(\beta_1\otimes\Theta))^\top\Sigma^{-1}(y-x\beta_{0}-b(\beta_1\otimes\theta))\\
&&~~~~~~~~~~~~~~~+(\Delta b(\tilde T)^\top \Theta \gamma)^\top+(b(t)^\top \Theta \gamma)^\top)\Big|dt\\
&\leq&  \exp(g_{10}+g_{11}\|y\|)\Big(|\Lambda_1(\tilde T)-\Lambda_2(\tilde T)|+\int_0^{\tau}|\Lambda_1(t)-\Lambda_2(t)|dt\Big)
\end{eqnarray*}
where $g_{10}$ and $g_{11}$ are two positive constants. Similarly,
\begin{eqnarray*}
&&|Q_3(t,O,\phi,\Lambda_1)-Q_3(t,O,\phi,\Lambda_2)|\\
&\leq&  \exp(g_{10}+g_{11}\|y\|)\Big(|\Lambda_1(\tilde T)-\Lambda_2(\tilde T)|+\int_0^{\tau}|\Lambda_1(t)-\Lambda_2(t)|dt\Big).
\end{eqnarray*}
On the other hand, there exist positive constants $g_{13},\cdots,g_{17}$ such that $|Q_1(t,O,\phi)|\leq \exp(g_{12}+g_{13}\|y\|)$, $Q_2(t,O,\phi,\Lambda_1)\leq (2\pi)^{k/2}$ and $Q_3(t,O,\phi,\Lambda_1)\geq \int_{\alpha}\exp(-\frac{1}{2}\alpha^\top\alpha-\exp(g_{14}+g_{15}\|\alpha\|+g_{16}\|y\|)B_0)d\alpha\geq g_{17}>0$. Therefore, by the mean-value theorem, we conclude that, for any $(t_1,O,\phi_1,\Lambda_1)$ and $(t_2,O,\phi_2,\Lambda_2)$ in $[0,\tau]\times \Phi \times \mathcal{Z}$,
\begin{eqnarray*}
&&|Q(t_1,O,\phi_1,\Lambda_1)-Q(t_2,O,\phi_2,\Lambda_2)|\\
&\leq&  \exp(g_{18}+g_{19}\|y\|)\Big(\|\phi_1-\phi_2\|+|\Lambda_1(\tilde T)-\Lambda_2(\tilde T)|+\int_0^{\tilde T}|\Lambda_1(t)-\Lambda_2(t)|dt+|t_1-t_2|\Big)
\end{eqnarray*}
holds for some positive constants $g_{18}$ and $g_{19}$.

According to Theorem 2.7.5 in Van Der Vaart and Wellner \cite{Van:1996}, the entropy number for the class $\mathcal{Z}$ satisfies $\log N_{[\cdot]}(\epsilon,\mathcal{Z},L_2(P))\leq K/\epsilon$, where $K$ is a constant. Thus, we can find $\exp(K/\epsilon)$ brackets, $\{[L_j,U_j]\}$, to cover the class $\mathcal{Z}$ such that, for each pair of $[L_j,U_j]$, $\|U_j-L_j\|_{L_2(P)}\leq \epsilon$. We can further find a partition of $[0,\tau]\times \Phi$, say $S_1 \cup S_2\cup \cdots,$ such that the number of partitions is of the order $(1/\epsilon)^{d+1}$ and for any $(t_1,\phi_1)$ and $(t_2,\phi_2)$ in the same partition, their Euclidean distance is less than $\epsilon$. Therefore, the partition $\{S_1,S_2,\cdots\}\times \{[L_j,U_j]\}$ bracket covers $[0,\tau]\times \Phi\times \mathcal{Z}$ and the total number of the partition is of order $(1/\epsilon)^{d+1}\exp(1/\epsilon)$. Furthermore, $\mathcal{F}$ has an $L_2(P)$-integrable covering function, which is equal to $O(e^{g_{18}+g_{19}\|y\|})$. From Theorem 2.5.6 in Van Der Vaart and Wellner \cite{Van:1996}, $\mathcal{F}$ is P-Donsker.

\medskip

In our model, the parameter $\psi=(\phi,\Lambda)\in \Psi = \{(\phi,\Lambda):\|\phi-\phi_0\|+\sup_{t\in[0,\tau]}|\Lambda(t)-\Lambda_0(t)|\leq \delta\}$ for a fixed small constant $\delta$. Define a set
$$\mathcal{H}=\{(\mathbf{h}_1,h_2):\|\mathbf{h}_1\|\leq 1, \|h_2\|_V\leq 1\}$$
where $\|h_2\|_V$ is the total variation of $h_2$ in $[0,\tau]$ defined as
$$\sup_{0=t_0\leq t_1<t_2<\cdots<t_m=\tau}\sum_{j=1}^m|h_2(t_j)-h_2(t_{j-1})|.$$
Moreover, let
\begin{eqnarray*}
S_n(\psi)(\mathbf{h}_1,h_2) &=& P_n\{l_{\phi}(\phi,\Lambda)^\top \mathbf{h}_1+l_{\Lambda}(\phi,\Lambda)[h_2]\},\\
S(\psi)(\mathbf{h}_1,h_2) &=& P\{l_{\phi}(\phi,\Lambda)^\top \mathbf{h}_1+l_{\Lambda}(\phi,\Lambda)[h_2]\},
\end{eqnarray*}
where $l_{\phi}(\phi,\Lambda)$ is the first derivative of the log-likelihood function from one subject, denoted by $l(O,\phi,\Lambda)$ with respect to $\phi$, and $l_{\Lambda}(\phi,\Lambda)[h_2]$ is the derivative of $l(O,\phi,\Lambda_{\epsilon})$ at $\epsilon=0$, where $\Lambda_{\epsilon}(t)=\int_0^t(1+\epsilon h_2(s))d\Lambda_0(s)$. Thus, it is easy to see that $S_n$ and $S$ are both maps from $\Psi$ to $l^{\infty}(\mathcal{H})$ and $\sqrt{n}(S_n(\psi)-S(\psi))$ is an empirical process in the space $l^{\infty}(\mathcal{H})$.

\vspace{1cm}
\begin{lemma}
\label{lemma2}
Under assumptions (C1)-(C5), \begin{eqnarray*}
\mathcal{G}&=&\{l_{\phi}(\phi,\Lambda)^\top \mathbf{h}_1+l_{\Lambda}(\phi,\Lambda)[h_2]-l_{\phi}(\phi_0,\Lambda_0)^\top \mathbf{h}_1-l_{\Lambda}(\phi_0,\Lambda_0)[h_2],\\
&&~~~~~~~\|\phi-\phi_0\|+\sup_{t\in[0,\tau]}|\Lambda(t)-\Lambda_0(t)|< \delta,(\mathbf{h}_1,h_2)\in\mathcal{H}\}
\end{eqnarray*} is P-Donsker.
\end{lemma}

\textbf{Proof of Lemma \ref{lemma2}}. If we let $(\mathbf{h}_1^{\Sigma},\mathbf{h}_1^{\alpha},\mathbf{h}_1^{\beta_0},\mathbf{h}_1^{\beta_1},\mathbf{h}_1^{\theta},\mathbf{h}_1^{\Theta},\mathbf{h}_1^{\eta},\mathbf{h}_1^{\gamma})$
be the corresponding components of $\mathbf{h}_1$ for the parameters $(\text{Vec}(\Sigma), \text{Vec}(D),\beta_0,\beta_1,\theta,\text{Vec}(\Theta),\eta,\gamma)$  respectively, then $l_{\phi}(\phi,\Lambda)^\top \mathbf{h}_1+l_{\Lambda}(\phi,\Lambda)[h_2]$ has the expression
\begin{eqnarray*}
&&\mu_1(O,\phi,\Lambda)^\top \mathbf{h}_1-\int_0^{\tilde T}\mu_2(t,O,\phi,\Lambda)^\top \mathbf{h}_1 d\Lambda(t)\\
&& +\Delta h_2(\tilde T)-\int_0^{\tilde T}\mu_3(t,O,\phi,\Lambda) h_2(t) d\Lambda(t)
\end{eqnarray*}
where
\begin{eqnarray*}
&&\mu_1(O,\phi,\Lambda)^\top \mathbf{h}_1\\
&=&\Big(\int_{\alpha}G(\alpha,O,\phi,\Lambda)d\alpha\Big)^{-1}\int_{\alpha}G(\alpha,O,\phi,\Lambda)\Big[\frac{\alpha^\top D^{-1}\mathcal{D}D^{-1}\alpha}{2}-\frac{1}{2}\text{Tr}(D^{-1}\mathcal{D})-\frac{1}{2}\text{Tr}(\Sigma^{-1}\mathcal{E})\\
&&+ \frac{1}{2}(y-x\beta_0-b(\beta_1\otimes \theta)-b(\beta_1\otimes \Theta)\alpha)^\top\Sigma^{-1}\mathcal{E}\Sigma^{-1}(y-x\beta_0-b(\beta_1\otimes \theta)-b(\beta_1\otimes \Theta)\alpha)\\
&&+x^\top\Sigma^{-1}(y-x\beta_0-b(\beta_1\otimes \theta)-b(\beta_1\otimes \Theta)\alpha)\mathbf{h}_1^{\beta_0}\\
&&+ (\beta_1\otimes \theta)_{\beta_1}^\top b^\top\Sigma^{-1}(y-x\beta_0-b(\beta_1\otimes \theta)-b(\beta_1\otimes \Theta)\alpha)\mathbf{h}_1^{\beta_1}\\
&&+(\beta_1\otimes \theta)_{\theta}^\top b^\top\Sigma^{-1}(y-x\beta_0-b(\beta_1\otimes \theta)-b(\beta_1\otimes \Theta)\alpha)\mathbf{h}_1^{\theta}\\
&&+\alpha^\top(\beta_1\otimes \Theta)_{\beta_1}^\top b^\top\Sigma^{-1}(y-x\beta_0-b(\beta_1\otimes \theta)-b(\beta_1\otimes \Theta)\alpha)\mathbf{h}_1^{\beta_1}\\
&&+\alpha^\top(\beta_1\otimes \Theta)_{\Theta}^\top b^\top\Sigma^{-1}(y-x\beta_0-b(\beta_1\otimes \theta)-b(\beta_1\otimes \Theta)\alpha)\mathbf{h}_1^{\Theta}\\
&&+\Delta(Z(\tilde T)^\top \mathbf{h}_1^{\eta}+b(\tilde T)^\top \theta\mathbf{h}_1^{\gamma} +b(\tilde T)^\top \mathbf{h}_1^{\theta}\gamma+b(\tilde T)^\top \Theta \alpha \mathbf{h}_1^{\gamma}+(b(\tilde T_i)^\top \Theta \alpha \gamma)_{\Theta}\mathbf{h}_1^{\Theta})\Big]d\alpha,
\end{eqnarray*}
\begin{eqnarray*}
&&\mu_2(t,O,\phi,\Lambda)^\top \mathbf{h}_1\\
&=&\Big(\int_{\alpha}G(\alpha,O,\phi,\Lambda)d\alpha\Big)^{-1}\int_{\alpha}G(\alpha,O,\phi,\Lambda)\Big[e^{Z(t)^\top \eta+b(t)^\top \theta\gamma +b(t)^\top \Theta \alpha \gamma}\\
&&\{Z(t)^\top \mathbf{h}_1^{\eta}+b(t)^\top \theta\mathbf{h}_1^{\gamma} +b(t)^\top \mathbf{h}_1^{\theta}\gamma+b(t)^\top \Theta \alpha \mathbf{h}_1^{\gamma}+(b(\tilde T_i)^\top \Theta \alpha \gamma)_{\Theta}\mathbf{h}_1^{\Theta}\}d\Lambda(t)\Big]d\alpha,
\end{eqnarray*}
and
\begin{eqnarray*}
&&\mu_3(t,O,\phi,\Lambda)\\
&=&\Big(\int_{\alpha}G(\alpha,O,\phi,\Lambda)d\alpha\Big)^{-1}\int_{\alpha}G(\alpha,O,\phi,\Lambda)\Big[e^{Z(t)^\top \eta+b(t)^\top \theta\gamma +b(t)^\top \Theta \alpha \gamma}\Big]d\alpha.
\end{eqnarray*}

For $j=1,2,3$, we denote $\nabla_{\phi}\mu_j$ and $\nabla_{\Lambda}\mu_j[d\Lambda]$ the derivatives of $\mu_j$ with respect to $\phi$ and $\Lambda$ along the path $\Lambda+\epsilon\delta\Lambda$. Similar to Lemma \ref{lemma1}, we can verify that $\nabla_{\Lambda}\mu_j[d\Lambda]=\int_{0}^t \mu_{j+3}(s,O,\phi,\Lambda)d\delta\Lambda(s)$ and that there exist two positive constants $r_1$ and $r_2$ such that
\begin{eqnarray*}
\sum_{j}\{|\mu_j|+|\nabla_{\phi}\mu_j|\}\leq e^{r_1+r_2\|by\|}.
\end{eqnarray*}
On the other hand, by the mean value theorem, we have that, for any $(\phi,\Lambda,\mathbf{h}_1,h_2)$ and $(\tilde\phi,\tilde\Lambda,\mathbf{\tilde h}_1,\tilde h_2)$ in $\Psi\times \mathcal{H}$,
\begin{eqnarray*}
&&l_{\phi}(\phi,\Lambda)^\top \mathbf{h}_1+l_{\Lambda}(\phi,\Lambda)[h_2]-l_{\phi}(\tilde\phi,\tilde\Lambda)^\top \mathbf{\tilde h}_1-l_{\Lambda}(\tilde\phi,\tilde\Lambda)[\tilde h_2]\\
&=& (\phi-\tilde\phi)^\top \nabla_{\phi}\mu_1(O,\phi^*,\Lambda^*)\mathbf{h}_1+\int_0^{\tilde T}\mu_4(t,O,\phi^*,\Lambda^*)^\top\mathbf{h}_1d(\Lambda-\tilde\Lambda)(t)\\
&&-\int_0^{\tilde T}(\phi-\tilde\phi)^\top \nabla_{\phi}\mu_2(t,O,\phi^*,\Lambda^*)\mathbf{h}_1d\Lambda(t)\\
&& -\int_0^{\tilde T}\int_0^{t}\mu_5(s,O,\phi^*,\Lambda^*)d(\Lambda-\tilde\Lambda)(s)\mathbf{h}_1d\Lambda(t)\\
&&-\int_0^{\tilde T}\mu_2(t,O,\phi^*,\Lambda^*)^\top\mathbf{h}_1d(\Lambda-\tilde\Lambda)(t)\\
&& -\int_0^{\tilde T}(\phi-\tilde\phi)^\top \nabla_{\phi}\mu_3(t,O,\phi^*,\Lambda^*){h}_2(t)d\Lambda(t)\\
&& -\int_0^{\tilde T}\int_0^{t}\mu_6(s,O,\phi^*,\Lambda^*)d(\Lambda-\tilde\Lambda)(s){h}_2(t)d\Lambda(t)\\
&& -\int_0^{\tilde T}\mu_3(t,O,\phi^*,\Lambda^*)^\top{h}_2(t)d(\Lambda-\tilde\Lambda)(t)\\
&& + \mu_1(O,\tilde\phi,\tilde\Lambda)^\top(\mathbf{h}_1-\mathbf{\tilde h}_1)-\int_0^{\tilde T}\mu_2(t,O,\tilde\phi,\tilde\Lambda)^\top(\mathbf{h}_1-\mathbf{\tilde h}_1)d\Lambda(t)\\
&& \Delta(h_2(\tilde T)-\tilde h_2(\tilde T))-\int_0^{\tilde T}\mu_3(t,O,\tilde\phi,\tilde\Lambda)^\top({h}_2(t)-{\tilde h}_2(t))d\Lambda(t)
\end{eqnarray*}
where $(\phi^*,\Lambda^*)$ is equal to $\epsilon^*(\phi,\Lambda)+(1-\epsilon^*)(\tilde\phi,\tilde\Lambda)$ for some $\epsilon^*\in[0,1]$. Thus,
\begin{eqnarray*}
&&|l_{\phi}(\phi,\Lambda)^\top \mathbf{h}_1+l_{\Lambda}(\phi,\Lambda)[h_2]-l_{\phi}(\tilde\phi,\tilde\Lambda)^\top \mathbf{\tilde h}_1-l_{\Lambda}(\tilde\phi,\tilde\Lambda)[\tilde h_2]|\\
&\leq& \exp(r_{1}+r_{2}\|y\|)\Big[\|\phi-\tilde\phi\| +\|\mathbf{h}_1-\mathbf{\tilde h}_1\|+|\Lambda(\tilde T)-\tilde\Lambda(\tilde T)|\\
&&+\int_0^{\tau}|\Lambda(t)-\tilde\Lambda(t)|[dt+d|h_2(t)|+d|\tilde h_2(t)|]\\
&& +|h_2(\tilde T)-\tilde h_2(\tilde T)|+\int_0^{\tau}|h_2(t)-\tilde h_2(t)|[d\Lambda_1(t)+d\Lambda_2(t)]
\end{eqnarray*}
where $d|h_2(t)|=dh_2^+(t)+dh_2^-(t)$ and $d|\tilde h_2(t)|=d\tilde h_2^+(t)+d\tilde h_2^-(t)$. Therefore, using the same arguments as in Lemma \ref{lemma1} and noting that $\log N_{[\cdot]}(\epsilon,\{h_2:\|h_2\|_{V}\leq B_1\},L_2(Q))\leq K/\epsilon$ for a constant $B_1$ and any probability measure $Q$ where $K$ is a constant (Theorem 2.7.5 of Van Der Vaart and Wellner \cite{Van:1996}), we have
\begin{eqnarray*}
\log N_{[\cdot]}(\epsilon,\mathcal{G},L_2(P))\leq O(\frac{1}{\epsilon} +\log(\epsilon)).
\end{eqnarray*}
Hence, $\mathcal{G}$ is P-Donsker.

Furthermore, we can show that
\begin{eqnarray*}
&&|l_{\phi}(\phi,\Lambda)^\top \mathbf{h}_1+l_{\Lambda}(\phi,\Lambda)[h_2]-l_{\phi}(\phi_0,\Lambda_0)^\top \mathbf{ h}_1-l_{\Lambda}(\phi_0,\Lambda_0)[ h_2]|\\
&\leq& \exp(r_{1}+r_{2}\|y\|)\Big[\|\phi-\phi_0\|+|\Lambda(\tilde T)-\Lambda_0(\tilde T)|\\
&&+\int_0^{\tau}|\Lambda(t)-\Lambda_0(t)|dt +\Big|\int_0^{\tilde T}\mu_3(t,O,\theta^*,\Lambda^*)h_2(t)d(\Lambda-d\Lambda_0)(t)\Big|\Big].
\end{eqnarray*}
If $\|\phi-\phi_0\|\rightarrow 0$ and $\sup_{t\in[0,\tau]}|\Lambda(t)-\Lambda_0(t)|\rightarrow 0$, the above expression converges to zero uniformly. Therefore,
\begin{eqnarray*}
&&\sup_{(\mathbf{h}_1,h_2)\in\mathcal{H}}P[l_{\phi}(\phi,\Lambda)^\top \mathbf{h}_1+l_{\Lambda}(\phi,\Lambda)[h_2]-l_{\phi}(\phi_0,\Lambda_0)^\top \mathbf{ h}_1-l_{\Lambda}(\phi_0,\Lambda_0)[ h_2]]^2\rightarrow 0.
\end{eqnarray*}

\vspace{1cm}
\begin{lemma}
\label{lemma3}
Under assumptions (C1)-(C5), we derived the expression of derivative operator $\nabla S_{\psi_0}$.
\end{lemma}

\textbf{Proof of Lemma \ref{lemma3}}. Since
\begin{eqnarray*}
&&l_{\phi}(\phi,\Lambda)^\top \mathbf{h}_1+l_{\Lambda}(\phi,\Lambda)[h_2]-l_{\phi}(\phi_0,\Lambda_0)^\top \mathbf{ h}_1-l_{\Lambda}(\phi_0,\Lambda_0)[ h_2]\\
&=& (\phi-\phi_0)^\top \{\nabla_{\phi}\mu_1(O,\phi^*,\Lambda^*)-\int_0^{\tilde T} \nabla_{\phi}\mu_2(t,O,\phi^*,\Lambda^*)d\Lambda_0(t)\}\mathbf{h}_1\\
&&+\mathbf{h}_1^\top \int_0^{\tau}I(t\leq \tilde T)\{\mu_4(t,O,\phi^*,\Lambda^*)^\top-\mu_2(t,O,\phi^*,\Lambda^*) \\ &&~~~-\mu_5(t,O,\phi^*,\Lambda^*)\int_t^{\tau}d\Lambda_0(s)\}d(\Lambda-\Lambda_0)(t)\\
&& -(\phi-\phi_0)^\top\int_0^{\tau} I(t\leq \tilde T)\nabla_{\phi}\mu_3(t,O,\phi^*,\Lambda^*){h}_2(t)d\Lambda_0(t)\\
&& -\int_0^{\tau}\{I(t\leq \tilde T)\mu_6(t,O,\phi^*,\Lambda^*)\int_t^{\tilde T}{h}_2(s)d\Lambda_0(s)\\
&&~~~~~~~+I(t\leq \tilde T)
\mu_3(t,O,\phi^*,\Lambda^*){h}_2(t)\}d(\Lambda-\Lambda_0)(t),
\end{eqnarray*}
then it is clear that
\begin{eqnarray*}
&&\nabla S_{\psi_0}(\phi-\phi_0,\Lambda-\Lambda_0)[\mathbf{h}_1,h_2]\\
&=& (\phi-\phi_0)^\top P\{\nabla_{\phi}\mu_1(O,\phi_0,\Lambda_0)-\int_0^{\tilde T} \nabla_{\phi}\mu_2(t,O,\phi_0,\Lambda_0)d\Lambda_0(t)\}\mathbf{h}_1\\
&&+\mathbf{h}_1^\top \int_0^{\tau}P\Big[I(t\leq \tilde T)\{\mu_4(t,O,\phi_0,\Lambda_0)-\mu_2(t,O,\phi_0,\Lambda_0)\\
 &&~~~~~~-\mu_5(t,O,\phi_0,\Lambda_0)\int_t^{\tilde T}d\Lambda_0(s)\}\Big]d(\Lambda-\Lambda_0)(t) \nonumber\\
 && -(\phi-\phi_0)^\top\int_0^{\tau} P\{I(t\leq \tilde T)\nabla_{\phi}\mu_3(t,O,\phi_0,\Lambda_0)\}{h}_2(t)d\Lambda_0(t)\\
&& -\int_0^{\tau}P\{I(t\leq \tilde T)\mu_6(t,O,\phi_0,\Lambda_0)\int_t^{\tilde T}{h}_2(s)d\Lambda_0(s)\\
&&~~~~~~~~~~+I(t\leq \tilde T)
\mu_3(t,O,\phi_0,\Lambda_0){h}_2(t)\}d(\Lambda-\Lambda_0)(t).
\end{eqnarray*}
Because, for $j=1,2,\cdots,6$,
\begin{eqnarray*}
&&\sup_{t\in[0,\tau]}\|\mu_j(t,O,\phi^*,\Lambda^*)-\mu_j(t,O,\phi_0,\Lambda_0)\|\\
&&\leq \exp(r_{3}+r_{4}\|y\|)(\|\phi^*-\phi_0\|+\sup_{t\in[0,\tau]}|\Lambda^*(t)-\Lambda_0(t)|),
\end{eqnarray*}
we have
\begin{eqnarray*}
&&l_{\phi}(\phi,\Lambda)^\top \mathbf{h}_1+l_{\Lambda}(\phi,\Lambda)[h_2]-l_{\phi}(\phi_0,\Lambda_0)^\top \mathbf{ h}_1-l_{\Lambda}(\phi_0,\Lambda_0)[ h_2]\\
&&=\nabla S_{\psi_0}(\phi-\phi_0,\Lambda-\Lambda_0)[\mathbf{h}_1,h_2]+o(\|\phi-\phi_0\|+\sup_{t\in[0,\tau]}|\Lambda(t)-\Lambda_0(t)|)(\|\mathbf{h}_1\|+\|h_2\|_V).
\end{eqnarray*}
Therefore, $S(\psi_0)$ is $Fr\acute{e}chet$ differentiable.

We can rewrite $\nabla S_{\psi_0}(\phi-\phi_0,\Lambda-\Lambda_0)[\mathbf{h}_1,h_2]$ as $(\phi-\phi_0)^\top \Omega_1[\mathbf{h}_1,h_2]+\int_0^{\tau}\Omega_2[\mathbf{h}_1,h_2]d(\Lambda-\Lambda_0)(t)$, where
\begin{eqnarray*}
\Omega_1[\mathbf{h}_1,h_2]
&=& \mathbf{h}_1^\top P\{\nabla_{\phi}\mu_1(O,\phi_0,\Lambda_0)-\int_0^{\tilde T} \nabla_{\phi}\mu_2(t,O,\phi_0,\Lambda_0)d\Lambda_0(t)\}\\
&& -\int_0^{\tau} P\{I(t\leq \tilde T)\nabla_{\phi}\mu_3(t,O,\phi_0,\Lambda_0)\}{h}_2(t)d\Lambda_0(t),
\end{eqnarray*}
and
\begin{eqnarray*}
\Omega_2[\mathbf{h}_1,h_2]
&=& \mathbf{h}_1^\top P\{I(t\leq \tilde T)\{\mu_4(t,O,\phi_0,\Lambda_0)-\mu_2(t,O,\phi_0,\Lambda_0)\\
 &&~~~~~-\mu_5(t,O,\phi_0,\Lambda_0)\int_t^{\tilde T}d\Lambda_0(s)\}\Big]\\
&& -P\{I(t\leq \tilde T)\mu_6(t,O,\phi_0,\Lambda_0)\int_t^{\tilde T}{h}_2(s)d\Lambda_0(s)\}\\
&&-P\{I(t\leq \tilde T)\mu_3(t,O,\phi_0,\Lambda_0)\}{h}_2(t).
\end{eqnarray*}
Then the operator $\Omega=(\Omega_1,\Omega_2)$ is the bounded linear operator from $R^d\times BV[0,\tau]$ to itself. Moreover, we note that $\Omega=A+(K_1,K_2)$, where $A(\mathbf{h}_1,h_2)=(\mathbf{h}_1,-PI(t\leq \tilde T)\mu_3(t,O,\phi_0,\Lambda_0)\}{h}_2(t))$, $K_1(\mathbf{h}_1,h_2)=\Omega_1[\mathbf{h}_1,h_2]-\mathbf{h}_1$ and
\begin{eqnarray*}
&&K_2(\mathbf{h}_1,h_2)\\
&=& \mathbf{h}_1^\top P\{I(t\leq \tilde T)\{\mu_4(t,O,\phi_0,\Lambda_0)-\mu_2(t,O,\phi_0,\Lambda_0)-\mu_5(t,O,\phi_0,\Lambda_0)\int_t^{\tilde T}d\Lambda_0(s)\}\}\\
&& -P\{I(t\leq \tilde T)\mu_6(t,O,\phi_0,\Lambda_0)\int_t^{\tilde T}{h}_2(s)d\Lambda_0(s)\}.
\end{eqnarray*}
Obviously, $A$ is invertible. Since $K_1$ maps into a finite-dimensional space, it is compact. The image of $K_2$ is a continuously differentiable function in $[0,\tau]$. According to the Arzela-Ascoli theorem, $K_2$ is a compact operator from $R^d\times BV[0,\tau]$ to $BV[0,\tau]$. Therefore, we conclude that $\Omega$ is the summation of an invertible operator and a compact operator.

\medskip

\vspace{1cm}
\textbf{Proof of Theorem 1}.

The maximum likelihood estimate $(\hat\phi,\hat\Lambda)$ exists following a similar discussion by Zeng and Cai \cite{Zeng:2005}. We will next show that, with probability one, $\hat\Lambda(\tau)$ is bounded as $n$ goes to infinity.

Define $\hat\xi=\log\hat\Lambda(\tau)$ and $\tilde\Lambda(t)=\frac{\hat\Lambda(t)}{e^{\hat\xi}}$, then $\tilde\Lambda(\tau)=1$.
We have $0\leq \frac{1}{n}l_n(\hat \phi,e^{\hat\xi}\tilde\Lambda)-\frac{1}{n}l_n(\hat \phi,\tilde\Lambda)$, where
\begin{eqnarray*}
&&\frac{1}{n}l_n(\hat\phi,\Lambda)\nonumber\\
 &=& \frac{1}{n}\sum_{i=1}^n\log\int_{\alpha} \frac{1}{(2\pi)^{n_iJ/2}|\hat\Sigma|^{1/2}}\nonumber\\
 &&\times\exp(-\frac{1}{2}(y_{i}-x_{i}\hat\beta_{0}-b_i(\hat\beta_1\otimes\hat\theta)-b_i(\hat\beta_1\otimes\hat\Theta)\alpha))^\top \hat\Sigma^{-1}\nonumber\\
&&~~~~~~~~~~~(y_{i}-x_{i}\hat\beta_{0}-b_i(\hat\beta_1\otimes\hat\theta)-b_i(\hat\beta_1\otimes\hat\Theta)\alpha))\nonumber\\
&&\times\Lambda\{\tilde T_i\}^{\Delta_i}\exp(\Delta_i(Z_i(\tilde T_i)^\top \hat\eta+b(\tilde T_i)^\top \hat\theta\hat\gamma  +b(\tilde T_i)^\top \hat\Theta \alpha \hat\gamma)\nonumber\\
&&~~~~-\int_{0}^{\tilde T_i}\exp(Z_i(t)^\top \hat\eta+b(t)^\top \hat\theta\hat\gamma +b(t)^\top \hat\Theta \alpha \hat\gamma)d\Lambda(t))\nonumber\\
&&~~\times\frac{1}{(2\pi)^{k/2}|\hat D|^{1/2}}\exp(-\frac{1}{2}\alpha^\top \hat D^{-1}\alpha)d\alpha\nonumber\\
&=& -\frac{1}{n}\sum_{i=1}^n\log(2\pi)^{n_iJ/2}|\hat\Sigma|^{1/2}-\frac{1}{n}\sum_{i=1}^n\log(2\pi)^{k/2}|\hat D|^{1/2}\nonumber\\
&&+\frac{1}{n}\sum_{i=1}^n\Delta_i(Z_i(\tilde T_i)^\top \hat\eta+b(\tilde T_i)^\top \hat\theta\hat\gamma)-\frac{1}{n}\sum_{i=1}^n\log|V_i|^{1/2} \nonumber\\
&& -\frac{1}{2n}\sum_{i=1}^n(y_{i}-x_{i}\hat\beta_{0}-b_i(\hat\beta_1\otimes\hat\theta))^\top\hat\Sigma^{-1}(y_{i}-x_{i}\hat\beta_{0}-b_i(\hat\beta_1\otimes\hat\theta))\nonumber\\
&&+\frac{1}{2n}\sum_{i=1}^n((\Delta_ib(\tilde T_i)^\top \hat\Theta \hat\gamma)^\top+ (b_i(\hat\beta_1\otimes\hat\Theta))^\top\hat\Sigma^{-1}(y_{i}-x_{i}\hat\beta_{0}-b_i(\hat\beta_1\otimes\hat\theta)))^\top V_i^{-1}((\Delta_ib(\tilde T_i)^\top \hat\Theta \hat\gamma)^\top\nonumber\\
&&~~~~~~~~~~~~~~~+ (b_i(\hat\beta_1\otimes\hat\Theta))^\top\hat\Sigma^{-1}(y_{i}-x_{i}\hat\beta_{0}-b_i(\hat\beta_1\otimes\hat\theta)))\nonumber\\
&&+\frac{1}{n}\sum_{i=1}^n\Big[\Delta_i\log\Lambda\{\tilde T_i\}+\log\int_{\alpha}\exp(-\frac{\alpha^\top\alpha}{2}-\int_{0}^{\tilde T_i}\exp( Q_{1i}(t,\alpha,\hat\phi))d\Lambda(t)))\Big]d\alpha,
\end{eqnarray*}
where $V_i=\hat D^{-1}+(b_i(\hat\beta_1\otimes\hat\Theta))^\top\hat\Sigma^{-1}b_i(\hat\beta_1\otimes\hat\Theta)$ and
\begin{eqnarray*}
&&Q_{1i}(t,\alpha,\hat\phi))\\
&=& Z_i(t)^\top \hat\eta+b(t)^\top \hat\theta\hat\gamma+b(t)^\top \hat\Theta\hat\gamma V_i^{-1/2}\alpha+b(t)^\top \hat\Theta\hat\gamma V_i^{-1}\\
&&\times((\Delta_ib(\tilde T_i)^\top \hat\Theta \hat\gamma)^\top+(b_i(\hat\beta_1\otimes\hat\Theta))^\top\hat\Sigma^{-1}(y_{i}-x_{i}\hat\beta_{0}-b_i(\hat\beta_1\otimes\hat\theta))).
\end{eqnarray*}
Thus, it follows that
\begin{eqnarray}
0&\leq& \frac{1}{n}\sum_{i=1}^n\Delta_i\hat\xi\nonumber\\
&&+\frac{1}{n}\sum_{i=1}^n\log\int_{\alpha}\exp\left(-\frac{\alpha^\top\alpha}{2}-e^{\hat\xi}\int_{0}^{\tilde T_i}\exp( Q_{1i}(t,\alpha,\hat\phi))d\tilde\Lambda(t)\right)d\alpha\nonumber\\
&&~~~~~~-\frac{1}{n}\sum_{i=1}^n\log\int_{\alpha}\exp\left(-\frac{\alpha^\top\alpha}{2}-\int_{0}^{\tilde T_i}\exp( Q_{1i}(t,\alpha,\hat\phi))d\tilde\Lambda(t)\right)d\alpha.
\end{eqnarray}

Based on the assumption, there exist some positive constants $C_1$, $C_2$, and $C_3$ such that $|Q_{1i}(t,\alpha,\hat\phi)|\leq C_1\|\alpha\|+C_2\|y_i\|+C_3$. Denote $\alpha_0$ the standard multivariate normal distribution. From the concavity of the logarithm function, we have
\begin{eqnarray*}
&&\log\int_{\alpha}\exp\left(-\frac{\alpha^\top\alpha}{2}-\int_{0}^{\tilde T_i}\exp( Q_{1i}(t,\alpha,\hat\phi))d\tilde\Lambda(t)\right)d\alpha\\
&=&(2\pi)^{k/2}\log E_{\alpha_0}[\exp(-\int_{0}^{\tilde T_i}\exp( Q_{1i}(t,\alpha,\hat\phi))d\tilde\Lambda(t))]\\
&\geq& (2\pi)^{k/2}\log E_{\alpha_0}[\exp(-\exp(C_1\|\alpha\|+C_2\|y_i\|+C_3))]\\
&\geq& (2\pi)^{k/2} E_{\alpha_0}[-\exp(C_1\|\alpha\|+C_2\|y_i\|+C_3)]\\
&=&-\exp(C_2\|y_i\|+C_4)
\end{eqnarray*}
where $C_4$ is another constant. Then
\begin{eqnarray*}
0&\leq& \frac{1}{n}\sum_{i=1}^n\Delta_i\hat\xi+\frac{1}{n}\sum_{i=1}^n\log\int_{\alpha}\exp\Big(-\frac{\alpha^\top\alpha}{2}-e^{\hat\xi}\int_{0}^{\tilde T_i}\exp( Q_{1i}(t,\alpha,\hat\phi))d\tilde\Lambda(t)\Big)d\alpha+C_5\\
&\leq& \frac{1}{n}\sum_{i=1}^n\Delta_i\hat\xi+\frac{1}{n}\sum_{i=1}^nI(\tilde T_i=\tau)\log\int_{\alpha}\exp\left(-\frac{\alpha^\top\alpha}{2}-e^{\hat\xi}\int_{0}^{\tilde T_i}\exp( Q_{1i}(t,\alpha,\hat\phi))d\tilde\Lambda(t)\right)d\alpha\\
&&+\frac{1}{n}\sum_{i=1}^nI(\tilde T_i\neq\tau)\log\int_{\alpha}\exp\left(-\frac{\alpha^\top\alpha}{2}\right)d\alpha+C_5\\
&\leq& \frac{1}{n}\sum_{i=1}^n\Delta_i\hat\xi+\frac{1}{n}\sum_{i=1}^nI(\tilde T_i=\tau)\\
&&~~~\times\log\int_{\alpha}\exp\left(-\frac{\alpha^\top\alpha}{2}-e^{\hat\xi}\int_{0}^{\tau}\exp( Q_{1i}(t,\alpha,\hat\phi))d\tilde\Lambda(t)\right)d\alpha+C_6
\end{eqnarray*}
where $C_6$ is a constant.

On the other hand, since for any $\Gamma\geq0$, and $x>0$, $\Gamma\log(1+x/\Gamma)\leq \Gamma \cdot x/\Gamma=x$, we have $e^{-x}\leq (1+x/\Gamma)^{-\Gamma}$. Therefore,
\begin{eqnarray*}
&&\int_{\alpha}\exp\left(-\frac{\alpha^\top\alpha}{2}-e^{\hat\xi}\int_{0}^{\tau}\exp( Q_{1i}(t,\alpha,\hat\phi))d\tilde\Lambda(t)\right)d\alpha\\
&\leq&\int_{\alpha}\exp\left(-\frac{\alpha^\top\alpha}{2}\right)\left(1+e^{\hat\xi}\int_{0}^{\tau}\exp( Q_{1i}(t,\alpha,\hat\phi))d\tilde\Lambda(t)/\Gamma\right)^{-\Gamma}d\alpha\\
&\leq&\int_{\alpha}\Gamma^{\Gamma}\exp\left(-\frac{\alpha^\top\alpha}{2}\right)\left(e^{\hat\xi}\int_{0}^{\tau}\exp( Q_{1i}(t,\alpha,\hat\phi))d\tilde\Lambda(t)\right)^{-\Gamma}d\alpha\\
&\leq&\int_{\alpha}\Gamma^{\Gamma}\exp\left(-\frac{\alpha^\top\alpha}{2}-\hat\xi\Gamma\right)\left(\int_{0}^{\tau}\exp( Q_{1i}(t,\alpha,\hat\phi))d\tilde\Lambda(t)\right)^{-\Gamma}d\alpha.
\end{eqnarray*}
Since $Q_{1i}(t,\alpha,\hat\phi)\geq -C_1\|\alpha\|-C_2\|y_i\|-C_3$
\begin{eqnarray*}
0&\leq&
C_6+\frac{1}{n}\sum_{i=1}^n\Delta_i\hat\xi+\frac{1}{n}\sum_{i=1}^nI(\tilde T_i=\tau)\\
&&\times\log\Big(\Gamma^{\Gamma}\exp(-\hat\xi\Gamma)\int_{\alpha}\exp\left(-\frac{\alpha^\top\alpha}{2}+C_1\Gamma\|\alpha\|+C_2\Gamma\|y_i\|+C_3\Gamma\Big)d\alpha\right)\\
&\leq&C_6+\frac{1}{n}\sum_{i=1}^n\Delta_i\hat\xi-\frac{\Gamma}{n}\sum_{i=1}^nI(\tilde T_i =\tau)\hat\xi+C_7(\Gamma)
\end{eqnarray*}
where $C_7(\Gamma)$ is a deterministic function of $\Gamma$, we can choose $\Gamma$ large enough such that $\frac{1}{n}\sum_{i=1}^n\Delta_i\hat\xi\leq \frac{\Gamma}{2n}\sum_{i=1}^nI(\tilde T_i =\tau)$. Therefore,
\begin{eqnarray*}
0&\leq&
C_6+C_7(\Gamma)-\frac{\Gamma}{2n}\sum_{i=1}^nI(\tilde T_i =\tau)\hat\xi,
\end{eqnarray*}
and we then have
\begin{eqnarray*}
\hat\xi&\leq&
\frac{2(C_6+C_7(\Gamma))}{\Gamma\sum_{i=1}^nI(\tilde T_i =\tau)/n}.
\end{eqnarray*}
Let $B_0=\exp(4(C_6+C_7(\Gamma))/\Gamma P(\tilde T =\tau))$. We conclude that $\hat\Lambda(\tau)=e^{\hat\xi}\leq B_0$.

\medskip

By Helly selection theorem, we can choose a subsequence of $\hat\Lambda$ such that $\hat\Lambda$ weakly converges to some right-continuous monotone function $\Lambda^{*}$ with probability 1. By choosing a sub-subsequence, we can further assume $\hat\phi \rightarrow \phi^*$ and then show that $\phi^*=\phi_0$ and $\Lambda^*=\Lambda_0$.

After differentiating $l_n(\phi,\Lambda)$ with respect to $\Lambda\{\tilde T_i\}$, we show that $\hat\Lambda$ satisfies the equation
\begin{eqnarray*}
\hat\Lambda\{\tilde T_j\}=\frac{\Delta_j}{nP_n[Q(t,O,\hat\phi,\hat\Lambda)I(\tilde T\geq t)]|_{t=\tilde T_j}}.
\end{eqnarray*}
Imitating the above equation, we can construct another function, denoted by $\bar \Lambda$, such that
\begin{eqnarray*}
\bar\Lambda\{\tilde T_j\}=\frac{\Delta_j}{nP_n[Q(t,O,\phi_0,\Lambda_0)I(\tilde T\geq t)]|_{t=\tilde T_j}}.
\end{eqnarray*}
Equivalently,
\begin{eqnarray*}
\bar\Lambda(t)=\frac{1}{n}\sum_{j=1}^n\frac{\Delta_jI(\tilde T_j\leq t)}{P_n[Q(t,O,\phi_0,\Lambda_0)I(\tilde T\geq t)]|_{t=\tilde T_j}}.
\end{eqnarray*}
We claim that $\bar\Lambda(t)$ uniformly converges to $\Lambda_0(t)$ in $[0,\tau]$. To prove this, note that
\begin{eqnarray*}
&&\sup_{t\in [0,\tau]}\Big|\bar\Lambda(t)-E\Big[\frac{\Delta I(\tilde T\leq t)}{P[Q(t,O,\phi_0,\Lambda_0)I(\tilde T\geq t)]|_{t=\tilde T}}\Big]  \Big|\\
&\leq&\sup_{t\in [0,\tau]}\Big|\frac{1}{n}\sum_{j=1}^n I(\tilde T_j\leq t)\Delta_j \Big[\frac{1}{P_n[Q(t,O,\phi_0,\Lambda_0)I(\tilde T\geq t)]}\\
&&~~~~~~~~~~~~-\frac{1}{P[Q(t,O,\phi_0,\Lambda_0)I(\tilde T\geq t)]}\Big]_{t=\tilde T_j} \Big|\\
&& +\sup_{t\in [0,\tau]}\Big|(P_n-P)\Big[\frac{\Delta I(\tilde T\leq t)}{P[Q(t,O,\phi_0,\Lambda_0)I(\tilde T\geq t)]|_{t=\tilde T}}\Big]  \Big|\\
&\leq&\sup_{t\in [0,\tau]}\Big|\frac{1}{P_n[Q(t,O,\phi_0,\Lambda_0)I(\tilde T\geq t)]}\\
&&-\frac{1}{P[Q(t,O,\phi_0,\Lambda_0)I(\tilde T\geq t)]} \Big|\\
&& +\sup_{t\in [0,\tau]}\Big|(P_n-P)\Big[\frac{\Delta I(\tilde T\leq t)}{P[Q(t,O,\phi_0,\Lambda_0)I(\tilde T\geq t)]|_{t=\tilde T}}\Big]  \Big|.\\
\end{eqnarray*}
According to Lemma \ref{lemma1}, $\{Q(t,O,\phi_0,\Lambda_0):t\in [0,\tau]\}$ is a bounded and Glivenko-Cantelli class. Since $\{I(\tilde T\geq t):t\in [0,\tau]\}$ is also a Glivenko-Cantelli class and the function $(f,g)\mapsto fg$ for any bounded two functions $f$ and $g$ is Lipschitz continuous, $\{I(\tilde T\geq t)Q(t,O,\phi_0,\Lambda_0):t\in [0,\tau]\}$ is a Glivenko-Cantelli class. Then we have that $\sup_{t\in [0,\tau]}|P_n(I(\tilde T\geq t)Q(t,O,\phi_0,\Lambda_0))-P(I(\tilde T\geq t)Q(t,O,\phi_0,\Lambda_0))|$ converges to 0. Moreover, from Lemma \ref{lemma1}, $P(I(\tilde T\geq t)Q(t,O,\phi_0,\Lambda_0)$ is larger than $P(I(\tilde T\geq t)\exp(-C_8-C_9\|y\|)$ for two constants $C_8$ and $C_9$, so is bounded from below. Thus, the first term on the right-hand side of the above inequality tends to zero. Additionally, since $\{\frac{ I(\tilde T\leq t)}{P[Q(t,O,\phi_0,\Lambda_0)I(\tilde T\geq t)]|_{t=\tilde T}}:t\in [0,\tau]\}$ is also a Glivenko-Cantelli class, the second term on the right-hand side of the above inequality vanishes as $n$ goes to infinity. Therefore, we conclude that $\bar\Lambda(t)$ uniformly converges to
\begin{eqnarray*}
E\Big[\frac{\Delta I(\tilde T\leq t)}{P[Q(t,O,\phi_0,\Lambda_0)I(\tilde T\geq t)]|_{t=\tilde T}}\Big].
\end{eqnarray*}
It is easy to verify that this limit is equal to $\Lambda_0(t)$. Thus, $\bar\Lambda$ uniformly converges to $\Lambda_0$ in $[0,\tau]$.

From the construction of $\bar\Lambda$, we have that
\begin{eqnarray}
\hat\Lambda(t)=\int_0^t \frac{P_n[Q(t,O,\phi_0,\Lambda_0)I(\tilde T\geq t)]}{P_n[Q(t,O,\hat\phi,\hat\Lambda)I(\tilde T\geq t)]} d\bar\Lambda(t),
\end{eqnarray}
so $\hat\Lambda(t)$ is absolutely continuous with respect to $\bar\Lambda(t)$. On the other hand, since $\{I(\tilde T\geq t):t\in [0,\tau]\}$ and $\mathcal{F}$ are both Glivenko-Cantelli classes, $\{I(\tilde T\geq t)Q(t,O,\phi,\Lambda):t\in[0,\tau],\phi\in\Phi,\Lambda\in \mathcal{Z},\Lambda(\tau)\leq B_0\}$ is also a Glivenko-Cantelli class. Thus
\begin{eqnarray*}
&&\sup_{t\in[0,\tau]}|(P_n-P)Q(t,O,\hat\phi,\hat\Lambda)I(\tilde T\geq t)|\\
&&~~~~~+\sup_{t\in[0,\tau]}|(P_n-P)Q(t,O,\phi_0,\Lambda_0)I(\tilde T\geq t)|\rightarrow 0~~~a.s.
\end{eqnarray*}
Using the bounded convergence theorem and the fact that $\hat\phi$ converges to $\phi^*$ and $\hat\Lambda$ weakly converges to $\Lambda^*$, $P Q(t,O,\hat\phi,\hat\Lambda)I(\tilde T\geq t)$ converges to $P Q(t,O,\phi^*,\Lambda^*)I(\tilde T\geq t)$ for each $t$; moreover, it is straightforward to check that the derivative of $P Q(t,O,\hat\phi,\hat\Lambda)I(\tilde T\geq t)$ with respect to $t$ is uniformly bounded, so $P Q(t,O,\hat\phi,\hat\Lambda)I(\tilde T\geq t)$ is equi-continuous with respect to $t$. Thus, by the Arzela-Ascoli theorem, uniformly in $t\in[0,\tau]$,
\begin{eqnarray*}
P Q(t,O,\hat\phi,\hat\Lambda)I(\tilde T\geq t)\rightarrow P Q(t,O,\phi^*,\Lambda^*)I(\tilde T\geq t).
\end{eqnarray*}
Then, it holds that, uniformly in $t\in [0,\tau]$,
\begin{eqnarray}
&&\frac{\hat\Lambda\{t\}}{\bar\Lambda\{t\}}=\frac{P_n Q(t,O,\phi_0,\Lambda_0)I(\tilde T\geq t)}{P_n Q(t,O,\hat\phi,\hat\Lambda)I(\tilde T\geq t)} \rightarrow \frac{P Q(t,O,\phi_0,\Lambda_0)I(\tilde T\geq t)}{P Q(t,O,\phi^*,\Lambda^*)I(\tilde T\geq t)}.
\end{eqnarray}
Furthermore,
\begin{eqnarray*}
\Lambda^*(t)=\int_0^t \frac{P[Q(t,O,\phi_0,\Lambda_0)I(\tilde T\geq t)]}{P[Q(t,O,\phi^*,\Lambda^*)I(\tilde T\geq t)]} d\Lambda_0(t).
\end{eqnarray*}
Since $\Lambda_0(t)$ is differentiable with respect to the Lebesgue measure, so is $\Lambda^*(t)$, we denote $\lambda^*(t)$ as the derivative of $\Lambda^*(t)$. Additionally, $\frac{\hat\Lambda\{t\}}{\bar\Lambda\{t\}}$ uniformly converges to $\frac{d\Lambda^*\{\tilde T\}}{d\Lambda_0\{\tilde T\}}=\frac{\lambda^*(\tilde T)}{\lambda_0(\tilde T)}$. It follows that $\hat\Lambda$ uniformly converges to $\Lambda^*$ since $\Lambda^*$ is continuous.

On the other hand,
\begin{eqnarray*}
&&\frac{1}{n}l_n(\hat \phi,\hat\Lambda)-\frac{1}{n}l_n(\phi_0,\bar\Lambda)=P_n\Big[\Delta\log\frac{\hat\Lambda\{\tilde T\}}{\bar\Lambda\{\tilde T\}}\Big]+P_n\Big[\log\frac{\int_{\alpha}G(\alpha,O,\hat\phi,\hat\Lambda)d\alpha}{\int_{\alpha}G(\alpha,O,\phi_0,\bar\Lambda)d\alpha}\Big]\geq 0.
\end{eqnarray*}
Using similar arguments as above, $\log\frac{\int_{\alpha}G(\alpha,O,\hat\phi,\hat\Lambda)d\alpha}{\int_{\alpha}G(\alpha,O,\phi_0,\bar\Lambda)d\alpha}$ belongs to a Glivenko-Cantelli class and
\begin{eqnarray*}
&&P\Big[\log\frac{\int_{\alpha}G(\alpha,O,\hat\phi,\hat\Lambda)d\alpha}{\int_{\alpha}G(\alpha,O,\phi_0,\bar\Lambda)d\alpha}\Big]\rightarrow  P\Big[\log\frac{\int_{\alpha}G(\alpha,O,\phi^*,\Lambda^*)d\alpha}{\int_{\alpha}G(\alpha,O,\phi_0,\Lambda_0)d\alpha}\Big].
\end{eqnarray*}
Therefore,
\begin{eqnarray*}
 P\Big[\log\frac{\lambda^*(\tilde T)^{\Delta}\int_{\alpha}G(\alpha,O,\phi^*,\Lambda^*)d\alpha}{\lambda_0(\tilde T)^{\Delta}\int_{\alpha}G(\alpha,O,\phi_0,\Lambda_0)d\alpha}\Big]\geq 0.
\end{eqnarray*}
Since the left-hand side of the inequality is the negative Kullback-Leibler information, it immediately follows that with probability one,
\begin{eqnarray}\label{identi}
&&\lambda^*(\tilde T)^{\Delta}\int_{\alpha}G(\alpha,O,\phi^*,\Lambda^*)d\alpha=\lambda_0(\tilde T)^{\Delta}\int_{\alpha}G(\alpha,O,\phi_0,\Lambda_0)d\alpha.
\end{eqnarray}

Setting $\Delta=0$ and $\tilde T=0$ in (\ref{identi}), we have
\begin{eqnarray}
&&\int_{\alpha}G(\alpha,O,\phi,\Lambda)d\alpha\nonumber\\
&=&\int_{\alpha}\frac{1}{(2\pi)^{n_iJ/2}|\Sigma|^{1/2}}\frac{1}{(2\pi)^{k/2}| D|^{1/2}}\nonumber\\
&&~~~~~~\times\exp\Big(-\frac{1}{2}(y_{i}-x_{i}\beta_{0}-b_i(\beta_1\otimes\theta)-b_i(\beta_1\otimes\Theta)\alpha))^\top \Sigma^{-1}\nonumber\\
&&~~~~~~~~~~~~\times(y_{i}-x_{i}\beta_{0}-b_i(\beta_1\otimes\theta)-b_i(\beta_1\otimes\Theta)\alpha)-\frac{1}{2}\alpha^\top  D^{-1}\alpha\Big)d\alpha\nonumber\\
&=&\frac{1}{(2\pi)^{n_iJ/2}|\Sigma|^{1/2}}\frac{|V_i|^{-1/2}}{| D|^{1/2}}\times\exp(-\frac{1}{2}(y_{i}-x_{i}\beta_{0}-b_i(\beta_1\otimes\theta))^\top \Sigma^{-1}(y_{i}-x_{i}\beta_{0}-b_i(\beta_1\otimes\theta))\nonumber\\
&&~~~~~~~~~+\frac{1}{2}[(b(\beta_1\otimes \Theta))^\top\Sigma^{-1}(y_{i}-x_{i}\beta_{0}-b_i(\beta_1\otimes\theta))]^\top\nonumber\\
 &&~~~~~~~~~~~~~\times V_i^{-1} [(b(\beta_1\otimes \Theta))^\top\Sigma^{-1}(y_{i}-x_{i}\beta_{0}-b_i(\beta_1\otimes\theta))].
\end{eqnarray}
By comparing the coefficients of $yy^\top$, $y$ and the constant term in the exponential parts in (\ref{identi}), we have that
\begin{eqnarray}
&&-\Sigma^{*-1}+\Sigma^{*-1} b(\beta_1^{*}\otimes \Theta^{*}) V^{*-1} (b(\beta_1^{*}\otimes \Theta^{*}))^\top \Sigma^{*-1} \label{yy}\\
&&= -\Sigma_0^{-1}+\Sigma_0^{-1} b(\beta_{10}\otimes \Theta_0) V_0^{-1} (b(\beta_{10}\otimes \Theta_0))^\top \Sigma_0^{*-1},\nonumber\\
&&x\beta_0^{*} + b(\beta_1^{*}\otimes\theta^{*}) = x\beta_{00} + b(\beta_{10}\otimes\theta_0)\label{y},\\
&&\frac{|V^*|^{-1/2}}{|\Sigma^*|^{1/2}|D^*|^{1/2}}=\frac{|V_0|^{-1/2}}{|\Sigma_0|^{1/2}|D_0|^{1/2}}.
\end{eqnarray}
Based on assumption (C4), (\ref{y}) indicates that $\beta_0^*=\beta_{00}$, $\beta_1^*=\beta_{10}$ and $\theta^*=\theta_0$.

Setting $\Delta=0$ in (\ref{identi}), we have
\begin{eqnarray*}
&&E_{\alpha}\exp(-\int_{0}^{\tilde T_i}\exp(Z_i(t)^\top \eta^*+b(t)^\top \theta^*\gamma^* +b(t)^\top \Theta^* \alpha \gamma^*)d\Lambda^*(t))\\
&&= E_{\alpha}\exp(-\int_{0}^{\tilde T_i}\exp(Z_i(t)^\top \eta_0+b(t)^\top \theta_0\gamma_0 +b(t)^\top \Theta_0 \alpha \gamma_0)d\Lambda_0(t)),
\end{eqnarray*}
where $\alpha$ follows a normal distribution and is the complete statistic in this normal family. Therefore,
\begin{eqnarray*}
&&\exp(-\int_{0}^{\tilde T_i}\exp(Z_i(t)^\top \eta^*+b(t)^\top \theta^*\gamma^* +b(t)^\top \Theta^* \alpha \gamma^*)d\Lambda^*(t))\\
&&= \exp(-\int_{0}^{\tilde T_i}\exp(Z_i(t)^\top \eta_0+b(t)^\top \theta_0\gamma_0 +b(t)^\top \Theta_0 \alpha \gamma_0)d\Lambda_0(t)).
\end{eqnarray*}
Equivalently
\begin{eqnarray*}
&&\exp(Z_i(t)^\top \eta^*+b(t)^\top \theta^*\gamma^* +b(t)^\top \Theta^* \alpha \gamma^*)\lambda^*(t)\\
&&=\exp(Z_i(t)^\top \eta_0+b(t)^\top \theta_0\gamma_0 +b(t)^\top \Theta_0 \alpha \gamma_0)\lambda_0(t).
\end{eqnarray*}
Assumptions (C4) and (C5) indicate that $\eta^*=\eta_0$, $\Theta^*=\Theta_0$ and $\gamma^*=\gamma_0$ and $\Lambda^*=\Lambda_0$.

Now consider $-\Sigma^{-1}+\Sigma^{-1} b(\beta_1\otimes \Theta) V^{-1} (b(\beta_1\otimes \Theta))^\top \Sigma^{-1}$ in (\ref{yy}). Let $W = \beta_1\otimes \Theta$. It can be expressed as $-\Sigma^{-1}+\Sigma^{-1} bW (D^{-1}+W^\top b^\top \Sigma^{-1}bW)^{-1} W^\top b^\top \Sigma^{-1}$. Taking the derivative of this matrix with respect to $b$, we have
\begin{eqnarray}\label{first derivative}
&&d(-\Sigma^{-1}+\Sigma^{-1} bW (D^{-1}+W^\top b^\top \Sigma^{-1}bW)^{-1} W^\top b^\top \Sigma^{-1})\nonumber\\
&=&\Sigma^{-1} dbW (D^{-1}+W^\top b^\top \Sigma^{-1}bW)^{-1} W^\top b^\top \Sigma^{-1}\nonumber\\
&&+\Sigma^{-1} bW (D^{-1}+W^\top b^\top \Sigma^{-1}bW)^{-1} W^\top db^\top \Sigma^{-1})\nonumber\\
&&+\Sigma^{-1} bW d(D^{-1}+W^\top b^\top \Sigma^{-1}bW)^{-1} W^\top b^\top \Sigma^{-1}\nonumber\\
&=:& I_1 + I_2+I_3.
\end{eqnarray}
For $I_1$, we have that
\begin{eqnarray*}
&&vec(I_1)\\
&=&vec(\Sigma^{-1} dbW (D^{-1}+W^\top b^\top \Sigma^{-1}bW)^{-1} W^\top b^\top \Sigma^{-1})\\
&=&(\Sigma^{-1}bW(D^{-1}+W^\top b^\top \Sigma^{-1}bW)^{-1}W^\top \otimes \Sigma^{-1}) vec(db).
\end{eqnarray*}
Similarly,
\begin{eqnarray*}
&&vec(I_2)\\
&=&vec(\Sigma^{-1} bW (D^{-1}+W^\top b^\top \Sigma^{-1}bW)^{-1} W^\top db^\top \Sigma^{-1}))\\
&=&K_{nJnJ}vec(\Sigma^{-1}dbW(D^{-1}+W^\top b^\top \Sigma^{-1}bW)^{-1}W^\top b^\top \Sigma^{-1}) \\
&=&K_{nJnJ}(\Sigma^{-1}bW(D^{-1}+W^\top b^\top \Sigma^{-1}bW)^{-1}W^\top \otimes \Sigma^{-1}) vec(db).
\end{eqnarray*}
Using the properties of the inverse function, we have
\begin{eqnarray*}
I_3
&=&\Sigma^{-1} bW d(D^{-1}+W^\top b^\top \Sigma^{-1}bW)^{-1} W^\top b^\top \Sigma^{-1}\\
&=&-\Sigma^{-1} bW (D^{-1}+W^\top b^\top \Sigma^{-1}bW)^{-1}d(W^\top b^\top \Sigma^{-1}bW)\\
&&~~~~~~~~~~~~(D^{-1}+W^\top b^\top \Sigma^{-1}bW)^{-1} W^\top b^\top \Sigma^{-1}\\
&=&-\Sigma^{-1} bW (D^{-1}+W^\top b^\top \Sigma^{-1}bW)^{-1}W^\top db^\top \Sigma^{-1}bW\\
&&~~~~~~~~~~~~(D^{-1}+W^\top b^\top \Sigma^{-1}bW)^{-1} W^\top b^\top \Sigma^{-1}\\
&&-\Sigma^{-1} bW (D^{-1}+W^\top b^\top \Sigma^{-1}bW)^{-1}W^\top b^\top \Sigma^{-1}dbW\\
&&~~~~~~~~~~~~(D^{-1}+W^\top b^\top \Sigma^{-1}bW)^{-1} W^\top b^\top \Sigma^{-1}.
\end{eqnarray*}
We vectorize $I_3$ and get
\begin{eqnarray*}
&&vec(I_3)\\
&=&-K_{nJnJ}vec(\Sigma^{-1} bW V^{-1}W^\top b^\top \Sigma^{-1}dbWV^{-1} W^\top b^\top \Sigma^{-1})\\
&&-vec(\Sigma^{-1} bW V^{-1}W^\top b^\top \Sigma^{-1}dbWV^{-1} W^\top b^\top \Sigma^{-1})\\
&=&-K_{nJnJ}(\Sigma^{-1}bWV^{-1}W^\top \otimes \Sigma^{-1}bWV^{-1}W^\top b^\top \Sigma^{-1}) vec(db)\\
&&-(\Sigma^{-1}bWV^{-1}W^\top \otimes \Sigma^{-1}bWV^{-1}W^\top b^\top \Sigma^{-1}) vec(db).
\end{eqnarray*}
Therefore,
\begin{eqnarray*}
&&dvec(-\Sigma^{-1}+\Sigma^{-1} bW (D^{-1}+W^\top b^\top \Sigma^{-1}bW)^{-1} W^\top b^\top \Sigma^{-1})\\
&=&(I+K_{nJnJ})(\Sigma^{-1}bWV^{-1}W^\top \otimes \Sigma^{-1}) vec(db)\\
&&-(I+K_{nJnJ})(\Sigma^{-1}bWV^{-1}W^\top \otimes \Sigma^{-1}bWV^{-1}W^\top b^\top \Sigma^{-1}) vec(db)\\
&=&(I+K_{nJnJ})(\Sigma^{-1}bWV^{-1}W^\top \otimes \Sigma^{-1}(I-bWV^{-1}W^\top b^\top \Sigma^{-1})) vec(db).
\end{eqnarray*}
We have that
\begin{eqnarray*}
&&\frac{dvec(-\Sigma^{-1}+\Sigma^{-1} bW (D^{-1}+W^\top b^\top \Sigma^{-1}bW)^{-1} W^\top b^\top \Sigma^{-1})}{dvec(b)}\\
&=&(I+K_{nJnJ})(\Sigma^{-1}bWV^{-1}W^\top \otimes \Sigma^{-1}(I-bWV^{-1}W^\top b^\top \Sigma^{-1})).
\end{eqnarray*}

Thus, $\Sigma^{*-1}bWV^{*-1}W^\top=\Sigma_0^{-1}bWV_0^{-1}W^\top$.
We further take the derivative with respect $b$ and have that
\begin{eqnarray*}
&&d(I+K_{nJnJ})(\Sigma^{-1}bWV^{-1}W^\top \otimes \Sigma^{-1}(I-bWV^{-1}W^\top b^\top \Sigma^{-1}))\\
&=&(I+K_{nJnJ})(d(\Sigma^{-1}bWV^{-1}W^\top) \otimes \Sigma^{-1}(I-bWV^{-1}W^\top b^\top \Sigma^{-1}))\\
&&+(I+K_{nJnJ})((\Sigma^{-1}bWV^{-1}W^\top) \otimes d(\Sigma^{-1}(I-bWV^{-1}W^\top b^\top \Sigma^{-1}))\\
&=:& J_1+J_2.
\end{eqnarray*}
Note that $d(\Sigma^{-1}(I-bWV^{-1}W^\top b^\top \Sigma^{-1})$ in $J_2$ is the same as the the function in (\ref{first derivative}), so we only consider $J_1$, for which we have
\begin{eqnarray*}
&&vec(J_1)\\
&=&vec((I+K_{nJnJ})(d(\Sigma^{-1}bWV^{-1}W^\top) \otimes \Sigma^{-1}(I-bWV^{-1}W^\top b^\top \Sigma^{-1})))\\
&=& (I\otimes(I+K_{nJnJ}))vec(d(\Sigma^{-1}bWV^{-1}W^\top) \otimes \Sigma^{-1}(I-bWV^{-1}W^\top b^\top \Sigma^{-1}))\\
&=& (I\otimes(I+K_{nJnJ}))(I_{qJ}\otimes K_{nJnJ}\otimes I_{nJ})\\
&&\times(vec(d(\Sigma^{-1}bWV^{-1}W^\top)) \otimes vec(\Sigma^{-1}(I-bWV^{-1}W^\top b^\top \Sigma^{-1}))).
\end{eqnarray*}
Since
\begin{eqnarray*}
vec(d(\Sigma^{-1}bWV^{-1}W^\top)) &=&vec(\Sigma^{-1}dbWV^{-1}W^\top)+vec(\Sigma^{-1}bWdV^{-1}W^\top)\\
&=&:J_1^{*}+J_2^{*},
\end{eqnarray*}
where
\begin{eqnarray*}
J_1^{*}=vec(\Sigma^{-1}dbWV^{-1}W^\top)=(WV^{-1}W^\top\otimes \Sigma^{-1}) vec(db),
\end{eqnarray*}
and
\begin{eqnarray*}
J_2^{*}&=&vec(\Sigma^{-1}bWdV^{-1}W^\top)\\
&=&-vec(\Sigma^{-1}bWV^{-1}(W^\top db^\top \Sigma^{-1}bW+W^\top b^\top \Sigma^{-1}dbW)V^{-1}W^\top)\\
&=&-K_{qJnJ}vec(WV^{-1}W^\top b^\top\Sigma^{-1}db WV^{-1}W^\top b^\top\Sigma^{-1})\\
&&~~~-vec(\Sigma^{-1}bWV^{-1}W^\top b^\top \Sigma^{-1}dbWV^{-1}W^\top)\\
&=&-K_{qJnJ}(\Sigma^{-1}bWV^{-1}W^\top \otimes WV^{-1}W^\top b^\top\Sigma^{-1}) vec(db)\\
&&-(WV^{-1}W^\top\otimes\Sigma^{-1}bWV^{-1}W^\top b^\top \Sigma^{-1})vec(db).
\end{eqnarray*}
we have
\begin{eqnarray*}
&&J_1^{*}+J_2^{*}\\
&=&-K_{qJnJ}(\Sigma^{-1}bWV^{-1}W^\top \otimes WV^{-1}W^\top b^\top\Sigma^{-1}) vec(db)\\
&&+(WV^{-1}W^\top\otimes\Sigma^{-1}(I-bWV^{-1}W^\top b^\top \Sigma^{-1}))vec(db).
\end{eqnarray*}
Hence, we prove that $V^{*}=V_0$, $\Sigma^{*}=\Sigma_0$, and $D^{*}=D_0$.

We conclude that, with probability 1, $\hat \phi$ converges to $\phi_0$ and $\hat\Lambda$ weakly converges to $\Lambda_0$
in $[0, \tau]$. However, since $\Lambda_0$ is continuous in $[0, \tau]$, the latter can be strengthened to uniform convergence; that is,
$\sup_{t\in[0,\tau ]} | \hat\Lambda(t)- \Lambda_0(t)|\rightarrow0$ almost surely. Hence, Theorem 1 is proved.

\vspace{1cm}
\textbf{Proof of Theorem 2}. The asymptotic properties for the estimators $(\hat\phi,\hat\Lambda)$ follow if we can verify the conditions in Van Der Vaart and Wellner \cite{Van:1996} for Theorem 3.3.1. which are given below. \\

Let $S_n$ and $S$ be random maps and a fixed map, respectively, from $\psi$ to a Banach space such that:\\
(a) $\sqrt{n}(S_n-S)(\hat \psi_n)-\sqrt{n}(S_n-S)(\psi_0)=o_p^*(1+\sqrt{n}\|\hat \psi_n-\psi_0\|)$,\\
(b) The sequence $\sqrt{n}(S_n-S)(\psi_0)$ converges in distribution to a tight random element $Z$,\\
(c) The function $\psi\rightarrow S(\psi)$ is $Fr\acute{e}chet$ differentiable at $\psi_0$ with a continuously invertible derivative $\nabla S_{\psi_0}$ (on its range),\\
(d) $S(\psi_0)=0$ and $\hat \psi_n$ satisfies $S_n(\hat \psi_n)=o_p^*(n^{-1/2})$ and converges in outer probability to $\psi_0$.\\
Then $\sqrt{n}(\hat \psi_n-\psi_0)\Rightarrow -\nabla S_{\psi_0}^{-1}Z$.

\medskip

According to Lemma \ref{lemma2}, the class
\begin{eqnarray*}
\mathcal{G}&=&\{l_{\phi}(\phi,\Lambda)^\top \mathbf{h}_1+l_{\Lambda}(\phi,\Lambda)[h_2]-l_{\phi}(\phi_0,\Lambda_0)^\top \mathbf{h}_1-l_{\Lambda}(\phi_0,\Lambda_0)[h_2],\\
&&~~~~~~~\|\phi-\phi_0\|+\sup_{t\in[0,\tau]}|\Lambda(t)-\Lambda_0(t)|< \delta,(\mathbf{h}_1,h_2)\in\mathcal{H}\}
\end{eqnarray*}
is P-Donsker. Moreover,
\begin{eqnarray*}
\sup_{(\mathbf{h}_1,h_2)\in\mathcal{H}}P[l_{\phi}(\phi,\Lambda)^\top \mathbf{h}_1+l_{\Lambda}(\phi,\Lambda)[h_2]-l_{\phi}(\phi_0,\Lambda_0)^\top \mathbf{h}_1-l_{\Lambda}(\phi_0,\Lambda_0)[h_2]]^2\rightarrow 0
\end{eqnarray*}
when $\|\phi-\phi_0\|+\sup_{t\in[0,\tau]}|\Lambda(t)-\Lambda_0(t)|\rightarrow 0$. Then (a) follows from Lemma 3.3.5 of Van Der Vaart and Wellner \cite{vaart1996weak}. By the Donsker theorem (Section 2.5 of Van Der Vaart and Wellner \cite{Van:1996}), (b) holds as a result of Lemma \ref{lemma2} and the convergence is defined in the metric space $l^{\infty}(\mathcal{H})$. Condition (d) is true since $(\hat \phi,\hat\Lambda)$ maximizes $P_nl(O,\phi,\Lambda)$, $(\phi_0,\Lambda_0)$ maximizes $Pl(O,\phi,\Lambda)$, and $(\hat \phi,\hat\Lambda)$ converges to $(\phi_0,\Lambda_0)$ from Theorem 1.

It remains to verify condition (c). The proof of the first half in (c) is tedious so we defer it to Lemma \ref{lemma3}. We only need to show that $\nabla S_{\psi_0}$ is continuously invertible on its range in $l^{\infty}(\mathcal{H})$. From Lemma \ref{lemma3}, $\nabla S_{\psi_0}$ can be written as follows: for any $(\phi_1,\Lambda_1)$ and $(\phi_2,\Lambda_2)$ in $\Psi$,
\begin{eqnarray*}
&&\nabla S_{\psi_0}(\phi_1-\phi_2,\Lambda_1-\Lambda_2)[\mathbf{h}_1,h_2]\\
&=&~~~~(\phi_1-\phi_2)^\top \Omega_1[\mathbf{h}_1,h_2]+\int_0^{\tau}\Omega_2[\mathbf{h}_1,h_2]d(\Lambda_1-\Lambda_2)(t),
\end{eqnarray*}
where both $\Omega_1$ and $\Omega_2$ are linear operators on $\mathcal{H}$ and $\Omega=(\Omega_1,\Omega_2)$ maps $\mathcal{H}\subset R^d \times BV[0,\tau]$, with $BV[0,\tau]$ containing all the functions with finite total variation in $[0,\tau]$. The explicit expressions of $\Omega_1$ and $\Omega_2$ are given in Lemma \ref{lemma3}. We can treat $(\phi_1-\phi_2,\Lambda_1-\Lambda_2)$ as an element in $l^{\infty}(\mathcal{H})$ via the following definition:
\begin{eqnarray*}
&&(\phi_1-\phi_2,\Lambda_1-\Lambda_2)[\mathbf{h}_1,h_2]=(\phi_1-\phi_2)^\top\mathbf{h}_1+\int_0^{\tau}h_2(t)d(\Lambda_1-\Lambda_2)(t)\\
&&~~~~~~~~~~~~~~~~~~~~~~~~~~~~~~~~~~~~~~~~~~~~~~~~~~~~~~~\forall (\mathbf{h}_1,h_2)\in R^d \times BV[0,\tau].
\end{eqnarray*}
Then $\nabla S_{\psi_0}$ can be expanded as a linear operator from $l^{\infty}(\mathcal{H})$ to itself. Therefore, if we can show that there exists some postive constant $\epsilon$ such that $\epsilon \mathcal{H}\subset \Omega(\mathcal{H})$, then for any $(\delta\phi,\delta\Lambda)\in l^{\infty}(\mathcal{H})$,
\begin{eqnarray*}
\|\nabla S_{\psi_0}(\delta\phi,\delta\Lambda)\|_{l^{\infty}(\mathcal{H})} &=& \sup_{(\mathbf{h}_1,h_2)\in \mathcal{H}}|\delta\phi^\top \Omega_1[\mathbf{h}_1,h_2]+\int_0^{\tau}\Omega_2[\mathbf{h}_1,h_2]d\delta\Lambda(t)|\\
&=& \|(\delta\phi,\delta\Lambda)\|_{l^{\infty}((\Omega(\mathcal{H}))}\geq \epsilon \|(\delta\phi,\delta\Lambda)\|_{l^{\infty}(\mathcal{H})}.
\end{eqnarray*}
Hence, $\nabla S_{\psi_0}$ is continuously invertible.

To prove $\epsilon \mathcal{H}\subset \Omega(\mathcal{H})$ for some $\epsilon$ is equivalent to showing that $\Omega$ is invertible. We note from Lemma \ref{lemma3} that $\Omega$ is the summation of an invertible operator and a compact operator. To prove the invertibility of $\Omega$, it is sufficient to verify that $\Omega$ is one to one: if $\Omega[\mathbf{h}_1,h_2]=0$, then by choosing $\phi_1-\phi_2=\tilde \epsilon \mathbf{h}_1$ and $\Lambda_1-\Lambda_2=\tilde \epsilon\int_0^{\tau}h_2d\Lambda_0$ for a small constant $\tilde \epsilon$, we obtain $\nabla S_{\psi_0}(\mathbf{h}_1,\int_0^{\tau}h_2d\Lambda_0)[\mathbf{h}_1,h_2]=0$. By the definition of $\nabla S_{\psi_0}$, we note that the left-hand side is the negative information matrix in the submodel $(\phi_0+ \epsilon \mathbf{h}_1, \Lambda_0+\epsilon\int_0^{\tau}h_2d\Lambda_0)$. Therefore, the score function along this submodel should be zero with probability one; that is, $l_{\phi}(\phi_0,\Lambda_0)^\top \mathbf{h}_1+l_{\Lambda}(\phi_0,\Lambda_0)[h_2]=0$. Therefore, if we let $(\mathbf{h}_1^{\Sigma},\mathbf{h}_1^{\alpha},\mathbf{h}_1^{\beta_0},\mathbf{h}_1^{\beta_1},\mathbf{h}_1^{\theta},\mathbf{h}_1^{\Theta},\mathbf{h}_1^{\eta},\mathbf{h}_1^{\gamma})$
be the corresponding components of $\mathbf{h}_1$ for the parameters $(\text{Vec}(\Sigma), \text{Vec}(D),\beta_0,\beta_1,\theta,\text{Vec}(\Theta),\eta,\gamma)$ respectively, and let $\mathcal{E}$, $\mathcal{D}$, $\mathcal{B}$ be the symmetric matrix such that $\text{Vec}(\mathcal{E})=\mathbf{h}_1^{\Sigma}$, $\text{Vec}(\mathcal{D})=\mathbf{h}_1^{\alpha}$, and $\text{Vec}(\mathcal{B})=\mathbf{h}_1^{\Theta}$, then with probability one,
\begin{eqnarray}\label{ze}
0&=&\int_{\alpha}G(\alpha,O,\phi_0,\Lambda_0)\Big[\frac{\alpha^\top D_0^{-1}\mathcal{D}D_0^{-1}\alpha}{2}-\frac{1}{2}\text{Tr}(D_0^{-1}\mathcal{D})-\frac{1}{2}\text{Tr}(\Sigma_0^{-1}\mathcal{E})\nonumber\\
&&+ \frac{1}{2}(y-x\beta_{00}-b(\beta_{10}\otimes \theta_0)-b(\beta_{10}\otimes \Theta_0)\alpha)^\top\Sigma_0^{-1}\mathcal{E}\Sigma_0^{-1}\nonumber\\
&&~~~~~~~~~~~~(y-x\beta_{00}-b(\beta_{10}\otimes \theta_0)-b(\beta_{10}\otimes \Theta_0)\alpha)\nonumber\\
&&+x^\top\Sigma_0^{-1}(y-x\beta_{00}-b(\beta_{10}\otimes \theta_0)-b(\beta_{10}\otimes \Theta_0)\alpha)\mathbf{h}_1^{\beta_0}\nonumber\\
&&+ (\beta_{10}\otimes \theta_0)_{\beta_1}^\top b^\top\Sigma_0^{-1}(y-x\beta_{00}-b(\beta_{10}\otimes \theta_0)-b(\beta_{10}\otimes \Theta_0)\alpha)\mathbf{h}_1^{\beta_1}\nonumber\\
&&+(\beta_{10}\otimes \theta_0)_{\theta}^\top b^\top\Sigma_0^{-1}(y-x\beta_{00}-b(\beta_{10}\otimes \theta_0)-b(\beta_{10}\otimes \Theta_0)\alpha)\mathbf{h}_1^{\theta}\nonumber\\
&&+\alpha^\top(\beta_{10}\otimes \Theta_0)_{\beta_1}^\top b^\top\Sigma_0^{-1}(y-x\beta_{00}-b(\beta_{10}\otimes \theta_0)-b(\beta_{10}\otimes \Theta_0)\alpha)\mathbf{h}_1^{\beta_1}\nonumber\\
&&+\alpha^\top(\beta_{10}\otimes \Theta_0)_{\Theta}^\top b^\top\Sigma_0^{-1}(y-x\beta_{00}-b(\beta_{10}\otimes \theta_0)-b(\beta_{10}\otimes \Theta_0)\alpha)\mathbf{h}_1^{\Theta}\nonumber\\
&&+\Delta(Z(\tilde T)^\top \mathbf{h}_1^{\eta}+b(\tilde T)^\top \theta_0\mathbf{h}_1^{\gamma} +b(\tilde T)^\top \mathbf{h}_1^{\theta}\gamma_0+b(\tilde T)^\top \Theta_0 \alpha \mathbf{h}_1^{\gamma}\nonumber\\
&&+(b(\tilde T_i)^\top \Theta_0 \alpha \gamma_0)_{\Theta}\mathbf{h}_1^{\Theta})-\int_{0}^{\tilde T}e^{Z(t)^\top \eta_0+b(t)^\top \theta_0\gamma_0 +b(t)^\top \Theta_0 \alpha \gamma_0}\nonumber\\
&&\times\{Z(t)^\top \mathbf{h}_1^{\eta}+b(t)^\top \theta_0\mathbf{h}_1^{\gamma} +b(t)^\top \mathbf{h}_1^{\theta}\gamma_0+b(t)^\top \Theta_0 \alpha \mathbf{h}_1^{\gamma}\nonumber\\
&&~~~~~~~~~~~~~~~~~~~~~~~~~~~~~~~+(b(\tilde T_i)^\top \Theta_0 \alpha \gamma_0)_{\Theta}\mathbf{h}_1^{\Theta})\}d\Lambda_0(t)\Big]d\alpha\nonumber\\
&&+\int_{\alpha}G(\alpha,O,\phi_0,\Lambda_0)\Big[\Delta h_2(\tilde T)\nonumber\\
&&~~~~~~~~~~~~~~~~~~~~~~-\int_{0}^{\tilde T}h_2(t)e^{Z(t)^\top \eta_0+b(t)^\top \theta_0\gamma_0 +b(t)^\top \Theta_0 \alpha \gamma_0}d\Lambda_0(t)\Big]d\alpha.
\end{eqnarray}
Setting $\Delta=0$ and $\tilde T=0$ in (\ref{ze}), we have
\begin{eqnarray*}
0&=&\int_{\alpha}G(\alpha,O,\phi_0,\Lambda_0)\Big[\frac{\alpha^\top D_0^{-1}\mathcal{D}D_0^{-1}\alpha}{2}-\frac{1}{2}\text{Tr}(D_0^{-1}\mathcal{D})-\frac{1}{2}\text{Tr}(\Sigma_0^{-1}\mathcal{E})\\
&&+ \frac{1}{2}(y-x\beta_{00}-b(\beta_{10}\otimes \theta_0)-b(\beta_{10}\otimes \Theta_0)\alpha)^\top\Sigma_0^{-1}\mathcal{E}\Sigma_0^{-1}\\
&&~~~~~~(y-x\beta_{00}-b(\beta_{10}\otimes \theta_0)-b(\beta_{10}\otimes \Theta_0)\alpha)\\
&&+x^\top\Sigma_0^{-1}(y-x\beta_{00}-b(\beta_{10}\otimes \theta_0)-b(\beta_{10}\otimes \Theta_0)\alpha)\mathbf{h}_1^{\beta_0}\\
&&+ (\beta_{10}\otimes \theta_0)_{\beta_1}^\top b^\top\Sigma_0^{-1}(y-x\beta_{00}-b(\beta_{10}\otimes \theta_0)-b(\beta_{10}\otimes \Theta_0)\alpha)\mathbf{h}_1^{\beta_1}\\
&&+(\beta_{10}\otimes \theta_0)_{\theta}^\top b^\top\Sigma_0^{-1}(y-x\beta_{00}-b(\beta_{10}\otimes \theta_0)-b(\beta_{10}\otimes \Theta_0)\alpha)\mathbf{h}_1^{\theta}\\
&&+\alpha^\top(\beta_{10}\otimes \Theta_0)_{\beta_1}^\top b^\top\Sigma_0^{-1}(y-x\beta_{00}-b(\beta_{10}\otimes \theta_0)-b(\beta_{10}\otimes \Theta_0)\alpha)\mathbf{h}_1^{\beta_1}\\
&&+\alpha^\top(\beta_{10}\otimes \Theta_0)_{\Theta}^\top b^\top\Sigma_0^{-1}(y-x\beta_{00}-b(\beta_{10}\otimes \theta_0)-b(\beta_{10}\otimes \Theta_0)\alpha)\mathbf{h}_1^{\Theta}\Big]d\alpha.
\end{eqnarray*}
Next, setting $\Delta=0$ in (\ref{ze}), we have
\begin{eqnarray*}
&&E_{\alpha}\Big[\exp(-\int_{0}^{\tilde T}e^{Z(t)^\top \eta_0+b(t)^\top \theta_0\gamma_0 +b(t)^\top \Theta_0 \alpha \gamma_0}d\Lambda_0(t))\\
&&\times \int_{0}^{\tilde T}e^{Z(t)^\top \eta_0+b(t)^\top \theta_0\gamma_0 +b(t)^\top \Theta_0 \alpha \gamma_0}\{Z(t)^\top \mathbf{h}_1^{\eta}+b(t)^\top \theta_0\mathbf{h}_1^{\gamma} +b(t)^\top \mathbf{h}_1^{\theta}\gamma_0\\
&&~~~~~~+b(t)^\top \Theta_0 \alpha \mathbf{h}_1^{\gamma}+(b(\tilde T_i)^\top \Theta_0 \alpha \gamma_0)_{\Theta}\mathbf{h}_1^{\Theta})+h_2(t)\}d\Lambda_0(t)\Big]=0,
\end{eqnarray*}
where $\alpha$ follows a normal distribution and is a complete statistic. Therefore,
\begin{eqnarray*}
&&\int_{0}^{\tilde T}e^{Z(t)^\top \eta_0+b(t)^\top \theta_0\gamma_0 +b(t)^\top \Theta_0 \alpha \gamma_0}\{Z(t)^\top \mathbf{h}_1^{\eta}+b(t)^\top \theta_0\mathbf{h}_1^{\gamma} +b(t)^\top \mathbf{h}_1^{\theta}\gamma_0\\
&&~~~+b(t)^\top \Theta_0 \alpha \mathbf{h}_1^{\gamma}+(b(\tilde T_i)^\top \Theta_0 \alpha \gamma_0)_{\Theta}\mathbf{h}_1^{\Theta})+h_2(t)\}d\Lambda_0(t)\Big]=0.
\end{eqnarray*}
Based on assumption (C4), we have that $\mathbf{h}_1^{\eta}=0$, $\mathbf{h}_1^{\gamma}=0$, $\mathbf{h}_1^{\theta}=0$, $\mathbf{h}_1^{\Theta}=0$, $h_2(t)=0$.

Since
\begin{eqnarray*}
0&=&\int_{\alpha}G(\alpha,O,\phi_0,\Lambda_0)\Big[\frac{\alpha^\top D_0^{-1}\mathcal{D}D_0^{-1}\alpha}{2}-\frac{1}{2}\text{Tr}(D_0^{-1}\mathcal{D})-\frac{1}{2}\text{Tr}(\Sigma_0^{-1}\mathcal{E})\\
&&+ \frac{1}{2}(y-x\beta_{00}-b(\beta_{10}\otimes \theta_0)-b(\beta_{10}\otimes \Theta_0)\alpha)^\top\Sigma_0^{-1}\mathcal{E}\Sigma_0^{-1}\\
&&~~~~~~(y-x\beta_{00}-b(\beta_{10}\otimes \theta_0)-b(\beta_{10}\otimes \Theta_0)\alpha)\\
&&+x^\top\Sigma_0^{-1}(y-x\beta_{00}-b(\beta_{10}\otimes \theta_0)-b(\beta_{10}\otimes \Theta_0)\alpha)\mathbf{h}_1^{\beta_0}\\
&&+ (\beta_{10}\otimes \theta_0)_{\beta_1}^\top b^\top\Sigma_0^{-1}(y-x\beta_{00}-b(\beta_{10}\otimes \theta_0)-b(\beta_{10}\otimes \Theta_0)\alpha)\mathbf{h}_1^{\beta_1}\\
&&+\alpha^\top(\beta_{10}\otimes \Theta_0)_{\beta_1}^\top b^\top\Sigma_0^{-1}(y-x\beta_{00}-b(\beta_{10}\otimes \theta_0)-b(\beta_{10}\otimes \Theta_0)\alpha)\mathbf{h}_1^{\beta_1}\Big]d\alpha,
\end{eqnarray*}
where $\alpha\sim N(m_{\alpha},V_{\alpha})$, $m_{\alpha}=V^{-1}(b(\beta_{10}\times\Theta_0))^\top\Sigma_0^{-1}(y-x\beta_{00}-b(\beta_{10}\otimes \theta_0))$, $V_{\alpha}=V^{-1}$, and $V=D_0^{-1}+(b(\beta_{10}\times\Theta_0))^\top\Sigma_0^{-1}(b(\beta_{10}\times\Theta_0))$, we have
\begin{eqnarray*}
&&\frac{m_{\alpha}^\top D_0^{-1}\mathcal{D}D_0^{-1}m_{\alpha}}{2}+\frac{1}{2}\text{Tr}(D_0^{-1}\mathcal{D}D_0^{-1}V_{\alpha})-\frac{1}{2}\text{Tr}(D_0^{-1}\mathcal{D})-\frac{1}{2}\text{Tr}(\Sigma_0^{-1}\mathcal{E})\\
&&+ \frac{1}{2}(y-x\beta_{00}-b(\beta_{10}\otimes \theta_0))^\top\Sigma_0^{-1}\mathcal{E}\Sigma_0^{-1}(y-x\beta_{00}-b(\beta_{10}\otimes \theta_0))\\
&& -(y-x\beta_{00}-b(\beta_{10}\otimes \theta_0))^\top\Sigma_0^{-1}\mathcal{E}\Sigma_0^{-1}b(\beta_{10}\otimes \Theta_0)m_{\alpha}\\
&&+ \frac{1}{2}m_{\alpha}^\top(b(\beta_{10}\otimes \Theta_0))^\top\Sigma_0^{-1}\mathcal{E}\Sigma_0^{-1}b(\beta_{10}\otimes \Theta_0)m_{\alpha}\\
&&+ \frac{1}{2}\text{Tr}((b(\beta_{10}\otimes \Theta_0))^\top\Sigma_0^{-1}\mathcal{E}\Sigma_0^{-1}b(\beta_{10}\otimes \Theta_0)V_{\alpha})\\
&&+x^\top\Sigma_0^{-1}(y-x\beta_{00}-b(\beta_{10}\otimes \theta_0)-b(\beta_{10}\otimes \Theta_0)m_{\alpha})\mathbf{h}_1^{\beta_0}\\
&&+ (\beta_{10}\otimes \theta_0)_{\beta_1}^\top b^\top\Sigma_0^{-1}(y-x\beta_{00}-b(\beta_{10}\otimes \theta_0)-b(\beta_{10}\otimes \Theta_0)m_{\alpha})\mathbf{h}_1^{\beta_1}\\
&&+m_{\alpha}^\top(\beta_{10}\otimes \Theta_0)_{\beta_1}^\top b^\top\Sigma_0^{-1}(y-x\beta_{00}-b(\beta_{10}\otimes \theta_0))\mathbf{h}_1^{\beta_1}\\
&&-m_{\alpha}^\top(\beta_{10}\otimes \Theta_0)_{\beta_1}^\top b^\top\Sigma_0^{-1}b(\beta_{10}\otimes \Theta_0)m_{\alpha})\mathbf{h}_1^{\beta_1}\\
&&-\text{Tr}((\beta_{10}\otimes \Theta_0)_{\beta_1}^\top b^\top\Sigma_0^{-1}b(\beta_{10}\otimes \Theta_0)\mathbf{h}_1^{\beta_1}V_{\alpha})=0.
\end{eqnarray*}
Examining the coefficient for $y-x\beta_{00}-b(\beta_{10}\otimes \theta_0)$ we know that $\mathbf{h}_1^{\beta_0}=0$, $\mathbf{h}_1^{\beta_1}=0$. The terms without $y-x\beta_{00}-b(\beta_{10}\otimes \theta_0)$ give
\begin{eqnarray}\label{cons}
&&\frac{1}{2}\text{Tr}(D_0^{-1}\mathcal{D}D_0^{-1}V_{\alpha})-\frac{1}{2}\text{Tr}(D_0^{-1}\mathcal{D})-\frac{1}{2}\text{Tr}(\Sigma_0^{-1}\mathcal{E})\nonumber\\ &&+\frac{1}{2}\text{Tr}((b(\beta_{10}\otimes \Theta_0))^\top\Sigma_0^{-1}\mathcal{E}\Sigma_0^{-1}b(\beta_{10}\otimes \Theta_0)V_{\alpha})=0.
\end{eqnarray}
Moreover, the coefficients for the quadratic term $(y-x\beta_{00}-b(\beta_{10}\otimes \theta_0))(y-x\beta_{00}-b(\beta_{10}\otimes \theta_0))^\top$ are equal to
\begin{eqnarray}
&&\frac{\Sigma_0^{-1}b(\beta_{10}\times\Theta_0)V^{-1} D_0^{-1}\mathcal{D}D_0^{-1}V^{-1}(b(\beta_{10}\times\Theta_0))^\top\Sigma_0^{-1}}{2}\nonumber\\
&&+ \frac{1}{2}\Sigma_0^{-1}\mathcal{E}\Sigma_0^{-1} -\Sigma_0^{-1}\mathcal{E}\Sigma_0^{-1}b(\beta_{10}\otimes \Theta_0)V^{-1}(b(\beta_{10}\times\Theta_0))^\top\Sigma_0^{-1}\nonumber\\
&&+ \frac{1}{2}\Sigma_0^{-1}b(\beta_{10}\times\Theta_0)V^{-1}(b(\beta_{10}\otimes \Theta_0))^\top\Sigma_0^{-1}\mathcal{E}\Sigma_0^{-1}\nonumber\\
&&\times b(\beta_{10}\otimes \Theta_0)V^{-1}(b(\beta_{10}\times\Theta_0))^\top\Sigma_0^{-1}=0.
\end{eqnarray}
Multiplying both sides by $(b(\beta_{10}\times\Theta_0))^\top$ from the left and by $b(\beta_{10}\times\Theta_0)$ from the  right gives
\begin{eqnarray*}
&&\frac{1}{2}(b(\beta_{10}\times\Theta_0))^\top\Sigma_0^{-1}b(\beta_{10}\times\Theta_0)V^{-1} D_0^{-1}\mathcal{D}D_0^{-1}V^{-1}\\
&&+ \frac{1}{2}(b(\beta_{10}\times\Theta_0))^\top\Sigma_0^{-1}\mathcal{E}\Sigma_0^{-1}b(\beta_{10}\times\Theta_0)((b(\beta_{10}\times\Theta_0))^\top\Sigma_0^{-1}b(\beta_{10}\times\Theta_0))^{-1}\\
&& -(b(\beta_{10}\times\Theta_0))^\top\Sigma_0^{-1}\mathcal{E}\Sigma_0^{-1}b(\beta_{10}\otimes \Theta_0)V^{-1}\\
&&+ \frac{1}{2}(b(\beta_{10}\times\Theta_0))^\top\Sigma_0^{-1}b(\beta_{10}\times\Theta_0)V^{-1}(b(\beta_{10}\otimes \Theta_0))^\top\\
&&~~~\times\Sigma_0^{-1}\mathcal{E}\Sigma_0^{-1}b(\beta_{10}\otimes \Theta_0)V^{-1}=0.
\end{eqnarray*}
Furthermore, multiplying the above equation by $VD_0$ from the right we have
\begin{eqnarray*}
&&\frac{1}{2}(b(\beta_{10}\times\Theta_0))^\top\Sigma_0^{-1}b(\beta_{10}\times\Theta_0)V^{-1} D_0^{-1}\mathcal{D}\\
&&+\frac{1}{2}(b(\beta_{10}\times\Theta_0))^\top\Sigma_0^{-1}\mathcal{E}\Sigma_0^{-1}b(\beta_{10}\times\Theta_0)((b(\beta_{10}\times\Theta_0))^\top\Sigma_0^{-1}b(\beta_{10}\times\Theta_0))^{-1}VD_0\\
&& -(b(\beta_{10}\times\Theta_0))^\top\Sigma_0^{-1}\mathcal{E}\Sigma_0^{-1}b(\beta_{10}\otimes \Theta_0)D_0\\
&&+ \frac{1}{2}(I-D_0^{-1}V_{\alpha})(b(\beta_{10}\otimes \Theta_0))^\top\Sigma_0^{-1}\mathcal{E}\Sigma_0^{-1}b(\beta_{10}\otimes \Theta_0)D_0=0,
\end{eqnarray*}
that is
\begin{eqnarray*}
&&\frac{1}{2}(b(\beta_{10}\times\Theta_0))^\top\Sigma_0^{-1}b(\beta_{10}\times\Theta_0)V^{-1} D_0^{-1}\mathcal{D}\\
&&+\frac{1}{2}(b(\beta_{10}\times\Theta_0))^\top\Sigma_0^{-1}\mathcal{E}\Sigma_0^{-1}b(\beta_{10}\times\Theta_0)((b(\beta_{10}\times\Theta_0))^\top\Sigma_0^{-1}b(\beta_{10}\times\Theta_0))^{-1}VD_0\\
&& -\frac{1}{2}(b(\beta_{10}\times\Theta_0))^\top\Sigma_0^{-1}\mathcal{E}\Sigma_0^{-1}b(\beta_{10}\otimes \Theta_0)D_0\\
&&- \frac{1}{2}D_0^{-1}V_{\alpha}(b(\beta_{10}\otimes \Theta_0))^\top\Sigma_0^{-1}\mathcal{E}\Sigma_0^{-1}b(\beta_{10}\otimes \Theta_0)D_0=0.
\end{eqnarray*}
Taking the trace of the matrix and using (\ref{cons}), we have
\begin{eqnarray*}
&&-\frac{1}{2}\text{Tr}(\Sigma_0^{-1}\mathcal{E})-\frac{1}{2}(b(\beta_{10}\times\Theta_0))^\top\Sigma_0^{-1}\mathcal{E}\Sigma_0^{-1}b(\beta_{10}\otimes \Theta_0)D_0\\
&&+\frac{1}{2}(b(\beta_{10}\times\Theta_0))^\top\Sigma_0^{-1}\mathcal{E}\Sigma_0^{-1}\\
&&~~~\times b(\beta_{10}\times\Theta_0)((b(\beta_{10}\times\Theta_0))^\top\Sigma_0^{-1}b(\beta_{10}\times\Theta_0))^{-1}VD_0=0.
\end{eqnarray*}
Hence, $\mathcal{E}=0$ and $\mathcal{D}=0$.

Now that conditions (a)-(d) have been proved, we conclude from Theorem 3.3.1 of Van Der Vaart and Wellner \cite{Van:1996} that $\sqrt{n}(\hat \phi-\phi_0,\hat \Lambda -\Lambda_0)$ weakly converges to a tight random element in $l^{\infty}(\mathcal{H})$. Moreover, we have
\begin{eqnarray}
&&\sqrt{n}\nabla S_{\psi_0}(\hat\phi-\phi_0,\hat\Lambda-\Lambda_0)[\mathbf{h}_1,h_2]\nonumber\\
&&=\sqrt{n}(P_n-P)(l_{\phi}(\phi_0,\Lambda_0)^\top \mathbf{h}_1+l_{\Lambda}(\phi_0,\Lambda_0)[h_2])+o_p(1),
\end{eqnarray}
where $o_p(1)$ is a random variable which converges to zero in probability in $l^{\infty}(\mathcal{H})$. Denoting $( \mathbf{\tilde h}_1,\tilde h_2)=\Omega^{-1}(\mathbf{h}_1,h_2)$, we then have
\begin{eqnarray}
&&\sqrt{n}\Big((\hat\phi-\phi_0)^\top \mathbf{h}_1+\int_0^{\tau}h_2(t)d(\hat\Lambda-\Lambda_0)(t)\Big)\nonumber\\
&&=\sqrt{n}(P_n-P)(l_{\phi}(\phi_0,\Lambda_0)^\top \mathbf{\tilde h}_1+l_{\Lambda}(\phi_0,\Lambda_0)[\tilde h_2])+o_p(1).
\end{eqnarray}
That is, $\sqrt{n}(\hat \phi-\phi_0,\hat \Lambda -\Lambda_0)$ weakly converges to a Gaussian process in $l^{\infty}(\mathcal{H})$.

\ \

\bibliographystyle{SageV}

\phantom{aaaa}